%% file: main.tex
\documentclass[letterpaper,twocolumn,10pt]{article}
\usepackage{usenix2019_v3}

\usepackage{tikz}
\usepackage{amsmath}
\usepackage{wrapfig}

\usepackage{filecontents}

\usepackage{tikz, filecontents, epsfig,endnotes,amsmath,algorithm,algorithmic,mathtools,longtable,multirow, amsmath,graphicx,verbatim,xcolor,cleveref,tablefootnote,amsfonts,caption}
\usepackage[frozencache=true,cachedir=minted-cache]{minted}
\usetikzlibrary{chains,shadows.blur,shapes.geometric,arrows}
\usepackage{booktabs}
\usepackage[normalem]{ulem}
\usepackage{array}
\usepackage{tabularray}
\usepackage[flushleft]{threeparttable}

\crefformat{section}{\S#2#1#3} %
\crefformat{subsection}{\S#2#1#3}
\crefformat{subsubsection}{\S#2#1#3}

\newcommand{\sysname}{{\sc AFLRun}}

\begin{document}

\date{}

\title{\Large \bf Toward Unbiased Multiple-Target Fuzzing with Path Diversity}

\author{
{\rm Huanyao Rong}\\
Indiana University Bloomington
\and
{\rm Wei You \thanks{Corresponding author.}}\\
\hspace{3.6mm}Renmin University of China\hspace{3.6mm}
\and
{\rm XiaoFeng Wang}\\
Indiana University Bloomington
\and
{\rm Tiaohao Mao}\\
Indiana University Bloomington
}

\maketitle

\begin{abstract}
\input{Abstract.tex}
\end{abstract}

\section{Introduction} \label{Introduction}

\input{Introduction.tex}

\section{Background} \label{Background}

\input{Background.tex}

\section{Motivation} \label{Motivation}

\input{Motivation.tex}

\section{Design Overview} \label{Overview}

\input{overview.tex}

\section{Target Path-Diversity Metric}

\input{Diversity.tex}

\section{Unbiased Energy Assignment}

\input{Weighted.tex}

\section{Evaluation}

\input{Evaluation.tex}

\section{Discussion}

\input{Discussion.tex}

\section{Related Work}
\input{Related.tex}

\section{Conclusion}
\input{Conclusion.tex}

\bibliographystyle{plain}
{\footnotesize{\bibliography{sample.bib}}}

\appendix
\input{Appendix.tex}

\end{document}

%% file: Abstract.tex
Directed fuzzing is an advanced software testing approach that systematically guides the fuzzing campaign toward user-defined target sites, enabling efficient discovery of vulnerabilities related to these sites.
However, we have observed that some complex vulnerabilities remain undetected by directed fuzzers even when the flawed target sites are frequently tested by the generated test cases, because triggering these bugs often requires the execution of additional code in related program locations.
Furthermore, when fuzzing multiple targets, the existing energy assignment in directed fuzzing lacks precision and does not ensure the fairness across targets, which leads to insufficient fuzzing effort spent on some deeper targets.

In this paper, we propose a novel directed fuzzing solution named \sysname{}, which features \textit{target path-diversity metric} and \textit{unbiased energy assignment}.
Firstly, we develop a new coverage metric by maintaining extra virgin map for each covered target to track the coverage status of seeds that hit the target. This approach enables the storage of waypoints that hit a target through interesting path into the corpus, thus enriching the path diversity for each target.
Additionally, we propose a corpus-level energy assignment strategy that ensures fairness for each target. \sysname{} starts with uniform target weight and propagates this weight to seeds to get a desired seed weight distribution. By assigning energy to each seed in the corpus according to such desired distribution, a precise and unbiased energy assignment can be achieved.

We built a prototype system and assessed its performance using a standard benchmark and several extensively fuzzed real-world applications. The evaluation results demonstrate that \sysname{} outperforms state-of-the-art fuzzers in terms of vulnerability detection, both in quantity and speed. Moreover, \sysname{} uncovers 29 previously unidentified vulnerabilities, including 8 CVEs, across four distinct programs.

%% file: Introduction.tex
Fuzzing is a testing technique that examines a program's behavior by subjecting it to a range of generated inputs, identifying any anomalous behaviors that arise and reporting inputs causing such anomalies. In recent years, coverage-guided fuzzing~\cite{Honggfuzz, libFuzzer, AFL, Fioraldi2020AFLC} has made a significant stride in uncovering software vulnerabilities.
However, coverage-guided fuzzing is designed to test all code regions within a program, but in certain security scenarios, such as static analysis report verification~\cite{DBLP:conf/icse/Christakis0W16} and patch testing~\cite{DBLP:conf/sigsoft/BohmeOR13}, the focus is on fuzzing specific code locations known as \emph{target sites}. As a result, directed fuzzing~\cite{Bhme2017DirectedGF}, building upon coverage-guided fuzzing, has emerged as an alternative, receiving a lot of research attention~\cite{Chen2018HawkeyeTA, sterlund2020ParmeSanSG, Zong2020FuzzGuardFO, Huang2022BEACONDG, Du2022WindrangerAD, Zheng2022FishFuzzTL, luo2022selectfuzz}.
The key idea of directed fuzzing is to approach and test the target sites by prioritizing the seeds (stored test cases) whose execution traces get to the vicinity of the target sites. This is achieved by evaluating the seeds using a distance oracle and assigning more \textit{energy} (which determines the effort spent on fuzzing a given seed) to those coming closer to the targets than others. In this way, the fuzzer is expected to quickly discover the test cases capable of reaching the target sites and activating the security flaws (the vulnerabilities) at these sites (i.e., triggering the crash).

\begin{listing}[!ht]
\begin{minted}[fontsize=\fontsize{7pt}{9pt}, xleftmargin=12pt,linenos]{c}
const char* get_string() {
  if (/*Easy Constraint*/) return "default";
  ... // Constaints solvable by genetic approach
  return get_input(); // Get attacker-controllable input
}
void foo() {
  char victim[128];
  const char* p = get_string();
  if (/*Hard Constraint*/)
    strcpy(victim, p); // Potential Overflow
}
\end{minted}
\caption{Example of Preconditioned Trigger}
\label{IntroExample}
\end{listing}

\vspace{2pt}\noindent\textbf{Limitations and challenges}. However, directed fuzzing today are impeded by the limitations of techniques, rendering it less effective than expected in discovering vulnerabilities. Specifically, as observed in our research, oftentimes, hitting a target site does not necessarily trigger the vulnerability it carries: that is, the fuzzing fails to produce a detectable crash. Instead, to activate the flaw, the execution trace needs to cover some other code locations, which could be far away from this target site. %
For example, in Listing~\ref{IntroExample}, the flaw at the target site (Line 10) can only be activated when the seed's execution goes through Line 4, instead of Line 2, since receiving a long input is the \textit{precondition} for overflowing the vulnerable buffer at the target site. We call such a scenario, in which a unique path needs to be covered before hitting the vulnerable target site for triggering such vulnerability at the target, \emph{preconditioned flaw activation} or \emph{PFA}. To sum up, PFA captures bugs requiring certain previous paths to trigger.
Another scenario is that at the target site only the \textit{root cause} of a security flaw is present while the flaw can only be activated in a different program location: as illustrated by the example in Listing~\ref{code:motivation}, the target site (Line 15-20) introduces an uninitialized pointer, which has only be dereferenced (thereby causing a crash) at Line 28 and 29. We call this scenario \emph{post-target activation} or \emph{PTA}, which captures bugs triggered at certain paths following the root cause.

Moreover, in the presence of multiple target sites, a directed fuzzer is supposed to generate test cases to fuzz these sites in an unbiased way. However, today's energy assignment strategy relies primarily on a scalar distance oracle that determines a seed's priority using average. As pointed by the prior work~\cite{Wang2020SoKTP, Liang2022MultipleTD, Zheng2022FishFuzzTL}, this strategy fails to take into account the dynamic status of each target (e.g., whether it is easier than others to reach), causing global optimum discrepancy, which could lead to \emph{a bias against certain targets or even favor the seed unable to reach any target}. As a result, deeper target sites may receive insufficient fuzzing effort, potentially leaving some vulnerabilities at these locations undiscovered.
Prior attempts to address the problem mostly focus on design of different distance oracles to mitigate the bias~\cite{Chen2018HawkeyeTA, Liang2022MultipleTD}. These approaches, however, are still less effective due to their continued use of aggregate information like average across all targets, thereby missing the individual information of each target, such as the extent to which the paths to a specific target have been covered by the current corpus. %
Even the most recent approach, FishFuzz~\cite{Zheng2022FishFuzzTL}, which utilizes a vector-based oracle instead of a distance oracle, still does not pay enough attention to the path diversity, ignoring the seed that hits a target visited frequently but is associated with the paths much less traversed than others.

\vspace{2pt}\noindent\textbf{Solutions}.
To address those limitations, we developed a new directed fuzzer by enhancing the greybox fuzzer with two innovative techniques. First, we introduce \emph{target path-diversity metric}. More specifically, our fuzzer, called \sysname{}, aims at generating the seeds not only reaching a given target site but covering as many paths involving the target as possible. The criticality of these paths in activating bugs is emphasized by the fact that enhancing path diversity can effectively address both PFA and PTA. For this purpose, \sysname{} maintains an extra virgin map for each target to record the coverage status of all seeds that hit the target. When deciding whether a mutated test case should be used to fuzz a program again (that is, storing the test case into the fuzzer's corpus as a seed), \sysname{} not only checks its execution trace with the virgin map of the coverage-guided fuzzer to evaluate its potential to improve the total code coverage, but also with our extra virgin maps to determine its capability of diversifying the execution paths involving the target site. 
It is important to note that the extra virgin map associated with the target captures coverage both before and after execution reaches the target. This approach enables the comprehensive capture of both PFA and PTA.

Second, we introduce \emph{unbiased energy assignment} to enable fair exploration and exploitation of each target. \sysname{} first assigns each target block (the basic block containing the target site) a weight, which takes the same value to ensure their fair chance to be visited. Then it propagates the weights from the targets to a set of \textit{critical blocks}, which include both the target blocks visited before and a set of \textit{critical boundary blocks}: that is, a visited block leading to a target via the inter-procedural control flow graph (ICFG), with none of other blocks (including the target) from the block to the target being covered before. For a given seed, our approach computes its energy (which determines its mutation priority) by adding the weights of the critical blocks it covers. In this way, \sysname{} always prioritizes the seeds with the highest potentials to get to a new target site and the seeds that cover any target sites, and also ensures each target to have a fair chance to be tested. When applied together with the targeted path-diversity metric, the energy assignment strategy can further improve the path diversity of each target.

We implemented \sysname{} on top of AFL++~\cite{Fioraldi2020AFLC} and AFLGo~\cite{Bhme2017DirectedGF}, and evaluated it on a standard benchmark Magma~\cite{Hazimeh2020MagmaAG} and real-world programs intensively fuzzed by the OSS-Fuzz project~\cite{OSSFuzz}. Our evaluation on the Magma benchmark shows that compared with state-of-the-art fuzzers (AFL++~\cite{Fioraldi2020AFLC}, AFLGo~\cite{Bhme2017DirectedGF}, Parmesan~\cite{sterlund2020ParmeSanSG}, FishFuzz~\cite{Zheng2022FishFuzzTL}, Hawkeye~\cite{Chen2018HawkeyeTA}, WindRanger~\cite{Du2022WindrangerAD} and MOpt~\cite{Lyu2019MOP}), \sysname{} achieves an average speedup of $168\%$, $109\%$, $235\%$, $183\%$, $147\%$, $157\%$ and $78\%$, and triggers $38\%$, $20\%$, $138\%$, $29\%$, $24\%$, $14\%$ and $50\%$ more vulnerabilities. In addition, \sysname{} has discovered 29 previously unknown vulnerabilities in real-world programs, with 8 CVE IDs assigned so far.

\vspace{2pt}\noindent\textbf{Contributions}. We have made the following contributions:

 \vspace{2pt}\noindent$\bullet$\textit{~New techniques}.  To the best of our knowledge, we present the first directed fuzzing technique that improves path diversity and fairness in energy assignment for each target at the same time, in a systematic way. Serving this purpose are two new techniques, including a new coverage metric that captures both PFA and PTA to improve the chance to activate the flaws related to target sites, and a new energy assignment strategy that ensures the target fairness and prioritizes testing of the new paths involving the targets. Both are found to be highly effective.

 \vspace{2pt}\noindent$\bullet$\textit{~Implementation, evaluation and findings}. We implemented a prototype system and released its code at \href{https://github.com/Mem2019/AFLRun}{https://github.com/Mem2019/AFLRun}. \sysname{} has discovered 29 zero-day vulnerabilities from the real-world software that have been intensively fuzzed before, which provides evidence to the efficacy of our techniques.

%% file: Background.tex
In this section, we introduce the basic concepts and overall workflow of coverage-guided fuzzing (\cref{CoverageGuidedFuzzing}) and directed fuzzing (\cref{DirectedFuzzing}).

\subsection{Coverage-guided Fuzzing} \label{CoverageGuidedFuzzing}

The workflow of the current coverage-guided fuzzing technique can be summarized as follows: (1) At the beginning of a fuzzing campaign, one or more inputs are provided as the initial seed corpus. (2) The fuzzer selects a seed from the corpus, which is then mutated to create a test case that is executed by the program under test (PUT); this mutation and execution process is repeated several times for each selected seed. (3) If the test case triggers any new behavior not yet triggered by the current corpus, it is stored as a seed for future mutation. (4) The fuzzer repeats steps 2-3 in an infinite loop. Since our approach focuses on improving energy assignment and seed storage by modifying steps 2 and 3 in AFL++~\cite{AFL, Fioraldi2020AFLC}, we will discuss how AFL++ implements these two steps in greater detail.

In AFL++, the fuzzer process and the PUT process share a segment of memory to record the execution path of the PUT, allowing the fuzzer to access this information. Each byte in the shared memory represents an edge connecting two basic blocks. The PUT is instrumented so that each time an edge is traversed during execution, the corresponding byte is incremented by one\footnote{AFL++ classifies value of each non-zero byte according to $8$ predefined ranges, each corresponding to a bit in the byte. Ultimately, the byte is set to the bit corresponding to the classified range. The comparison between the execution trace and the virgin map is performed at bit level.}.
In the shared memory, the fuzzer obtains a bitmap that abstractly represents the execution path of a single run. In this paper, we refer to this bitmap as the \emph{execution trace}. Additionally, AFL++ maintains a global bitmap to record all bits covered by at least one execution trace of seeds in the corpus so far. We refer to this bitmap as the \emph{virgin map} in this paper. Specifically, the virgin map is an array with the same length as the execution trace, initialized with all bits set to $1$. When a new seed covers a new bit, the corresponding bit in the virgin map is set to $0$. Each time a mutated test case is executed, AFL++ compares the generated execution trace with the virgin map. If the execution trace covers a new bit that has not been covered by any previous seeds' execution traces, the test case will be added to the corpus as a seed.
By incrementally adding new seeds into corpus, total coverage of the PUT can be improved.

In addition, AFL++ maintains a record of the best seed for each edge using an array, based on metrics such as the seed's length and execution time. We refer to this array as the \emph{top-rated array} in this paper. When a new seed is added to the corpus, AFL++ attempts to update the top-rated array based on the seed's edge coverage and metrics. If any updates occur, a process called \emph{queue culling} selects a subset of seeds from the corpus to be marked as favored, and these favored seeds are chosen for fuzzing more frequently than the others.

\subsection{Directed Fuzzing} \label{DirectedFuzzing}

Directed fuzzing, built upon coverage-based fuzzing, focuses on testing a specific set of user-defined target sites rather than all code regions. The main concept involves using a seed oracle (i.e., dynamic information gathered by executing the PUT with the seed as input) to determine the proximity between the seed and the target sites. This oracle is then used to assign energy for preferentially testing these sites. The energy assigned to a seed is defined by the number of times the fuzzer is going to mutate the seed and execute the PUT using the mutated test case as input.

The seed oracle is commonly referred to as the \emph{distance} in the context of directed fuzzing. To achieve such an oracle, both static analysis and dynamic information are required. Static analysis of the PUT computes the distance from each code location (e.g., basic block) to all target sites, typically averaged using the harmonic mean. During a fuzzing campaign, each time the PUT is executed with a particular seed as input, the seed's distance value to target sites (i.e., seed oracle) can be obtained by calculating the arithmetic mean of the distance values of the relevant code locations covered during the execution.

With this seed oracle, directed fuzzing assigns energy to each seed such that seeds with lower distance values receive higher amounts of energy. By adjusting energy assignment to favor seeds closer to the target sites, the fuzzing campaign focuses on testing these specific sites.

%% file: Motivation.tex
In this section, we present a PTA vulnerability as our motivating example and discuss the directed fuzzing scenario intended to discover such a vulnerability. (\cref{ExampleVuln}) We also examine how current directed fuzzing approaches fail to find it, while \sysname{} is capable of successfully detecting it. (\cref{LimitationsSolutions})

\subsection{Vulnerability and Fuzzing Scenario} \label{ExampleVuln}

We use the discovery of a zero-day vulnerability (CVE-2023-25588) in \texttt{binutils} as an example to demonstrate the limitations of existing techniques and motivate the ideas of \sysname{}.
Listing~\ref{code:motivation} shows the simplified vulnerable code snippet. The \texttt{dump\_bfd} function invokes the \texttt{bfd\_get\_synthetic\_symtab} function pointer (Line 10) to create synthetic symbols for indirect symbols and invokes the \texttt{disassemble\_data} function (Line 11) to disassemble the contents of an object file. When processing a Mach-O executable, the function pointer points to the \texttt{bfd\_mach\_o\_get\_synthetic\_symtab} function (Line 15-21), in which an array of the \texttt{asymbol} structure is allocated without initialization for its \texttt{the\_bfd} pointer field. A pointer to the allocated array is stored in the global variable \texttt{synthsyms}. The \texttt{disassemble\_data} function performs complex logic and finally calls the \texttt{compare\_symbols} function to sort symbols. During comparison, the \texttt{the\_bfd} field of the elements stored in the \texttt{synthsyms} array will be dereferenced, resulting in access of uninitialized pointer.

\input{cg_moti.tex}
\input{motivation_code.tex}

Figure~\ref{fig:motivation} illustrates the partial call graph involving the critical functions needed to trigger the vulnerability. In the figure, the solid lines represent normal function calls linking caller and callee, the dashed line indicates a indirect call unable to be identified by static distance computation, and the zigzag line shows multiple normal calls with omitted intermediate nodes. As depicted, in order to trigger the vulnerability at the location of dereference (Line 28 in \texttt{compare\_symbols}), a test case must cover this location after executing the location of allocation (Line 17 in \texttt{bfd\_mach\_o\_get\_synthetic\_symtab}, the root cause location of the vulnerability). It is important to note that these two locations are distant from each other in the code base.

Consider a security scenario that utilizes directed fuzzing to verify a report from static analysis. An aggressive, yet comprehensive and scalable static checker might identify potential risks if it detects a function that allocates a piece of heap memory without initializing the content within the function scope.
When employing such a checker on \texttt{binutils}, it reports multiple program locations, which are treated as target sites for directed fuzzing. The root cause of this vulnerability, located at Line 17, is one of these target sites.
Additionally, there are other false positive target sites, including Line 34 at \texttt{xmalloc}, which is simply a wrapper for the \texttt{malloc} function. Figure \ref{fig:motivation} also includes some functions that invoke \texttt{xmalloc}. For simplicity, we only show two functions that call \texttt{xmalloc}, as it is a utility commonly used by numerous functions.
Thus, an ideal directed fuzzer should be able to discover the vulnerability by generating a test case that both creates and dereferences the uninitialized variable, despite the interference from false positive target sites like Line 34.
Besides, it is worth noting that these target locations conform to the format used by AFLGo \cite{Bhme2017DirectedGF}, instead of targets with explicit dependency information required by some works \cite{Nguyen2020BinarylevelDF,Lee2021ConstraintguidedDG}.

\subsection{Limitations and Solutions} \label{LimitationsSolutions}

However, the current state-of-the-art directed fuzzer struggles to efficiently trigger this bug under the scenario we described earlier, even if the flawed target site at \texttt{bfd\_mach\_o\_get\_synthetic\_symtab} is already covered by the corpus during the fuzzing campaign. We outline two limitations that contribute to this issue and explain how our approaches address them.

First, current directed fuzzers still rely on the coverage metric of coverage-guided fuzzing, which is designed to improve the total code coverage of the corpus. Specifically, once a constraint is solved, the edge of the constraint will be marked as non-virgin in the virgin map, meaning that subsequent test cases that cover the same edge will not be stored in the corpus.
In this example vulnerability, if the constraints on the path to Line 28 (i.e., the dereference of \texttt{the\_bfd}) are solved with a seed that cannot cover Line 17 (i.e., the location that creates the uninitialized pointer), the edges of these constraints are no longer considered virgin.
Consequently, if a test case that covers the target site of uninitialized pointer allocation later solves any of these constraints, it will not be stored in the corpus. As a result, seeds that cover this target site cannot make progress on these constraints by taking advantage of the genetic approach employed by the greybox fuzzer, which ultimately restricts the subsequent fuzzing campaign from discovering this potential crash.

We have observed that, in order to trigger this bug, path diversity of the target site with the uninitialized pointer allocation is crucial.
In other words, we aim to maximize the total code coverage given the target site Line 17 is covered, in order to increase the likelihood of executing the code location that accesses the uninitialized pointer.
Our approach, by maintaining an additional virgin map for this target site, does store test cases that improve path diversity of this target site in the corpus, even if the original coverage metric of AFL++ would deem the test case uninteresting.

Second, the energy assignment in existing directed fuzzing, which relies on the scalar distance oracle, may be biased toward certain targets, resulting in inadequate fuzzing efforts spent on the vulnerable target site.
In this particular example scenario, we demonstrate the bias toward the target Line 34.
For simplicity, we only consider these two target sites and part of the program illustrated in Figure \ref{fig:motivation}, and assume there are only two categories of seeds in the corpus: $S_a$ and $S_b$. Seeds in $S_a$ can visit target site Line 17, while those in $S_b$ cannot. Both categories of seeds are able to cover \texttt{xmalloc} through \texttt{slurp\_symtab} and \texttt{disassemble\_data}.
Since Line 17 is an uncommonly covered target, $S_a$ makes up only a small proportion of the corpus.
To avoid loss of generality, we calculate the seed distance values using a simplified method that approximately represents current directed fuzzing approaches~\cite{Bhme2017DirectedGF, sterlund2020ParmeSanSG, Chen2018HawkeyeTA}: the distance value of each block is computed by the average of minimum distances to all targets that it can reach via the graph.
Using the distance value of each function in Figure \ref{fig:motivation}, we can calculate the distance values of $S_a$ and $S_b$ as $\frac{2+1+1+0+0}{5}=0.8$ and $\frac{2+1+1+0}{4}=1$, respectively.
It is worth noting that covering Line 17 only contributes one additional term and does not significantly influence the average scalar value. As a result, the final energy assigned to each seed in $S_a$ and $S_b$ does not differ greatly. However, since the size of $S_a$ is small, the total energy assigned to seeds in $S_a$ is also low, causing the fuzzing campaign to lose directness toward target site Line 17 and become biased toward Line 34.

Instead, as we do not know which target site contains the vulnerability beforehand, the importance of each target is unknown a priori. Therefore, a best scheme is to treat each target (e.g., Line 17 and Line 34) equally during the fuzzing campaign. Our approach achieves this by assigning a uniform weight to each target and propagating such weight to seeds for energy assignment.
To be specific, the weight assigned to $S_a$ is $\frac{1}{|S_a|} + \frac{1}{|S_a|+|S_b|}$, and the weight assigned to $S_b$ is $\frac{1}{|S_a|+|S_b|}$.\footnote{Here, we assume that the seeds in \(|S_a|\) and \(|S_b|\) have equal importance; however, in reality, the energy assignment among seeds in such set is not uniform. See \cref{EnergyAssignment} for more details.} Since $|S_a|$ is a much smaller value than $|S_b|$, the weight assigned to each seed of $S_a$ is significantly greater than that of $S_b$.
Therefore, even though $S_a$ comprises only a small set of seeds in the corpus, Line 17 will still be assigned an equal amount of weight as the others, ensuring that it receives enough energy to trigger the bug.

%% file: cg_moti.tex
\begin{figure}
\centering
\usetikzlibrary{shapes.geometric}
\usetikzlibrary{decorations.pathmorphing}
\tikzstyle{box}=[rectangle,text centered,draw=black]
\tikzstyle{round}=[ellipse,text centered,draw=black]
\begin{tikzpicture}[decoration=zigzag]
    \node[box,label={[red]above:2}](root){dump\_bfd};
    \node[box,below of=root,xshift=-3cm,yshift=-0.25cm,label={[red]above:1}](leaf11){slurp\_symtab};
    \node[box,below of=root,xshift=-0.5cm,yshift=-0.4cm,align=center,label={[red]above:0}](leaf12){bfd\_mach\_o\_ \\ get\_synthetic\_ \\ symtab};
    \node[box,below of=root,xshift=2.5cm,yshift=-0.25cm,label={[red]above:1}](leaf13){disassemble\_data};
    \node[box,below of=leaf13,xshift=0.25cm,yshift=-0.1cm](leaf22){compare\_symbols};
    \node[box,below of=leaf12,xshift=0.8cm,yshift=-0.45cm,label={[red]above:0}](l3){xmalloc};
    \draw[->](root.west) to (leaf11);
    \draw[->,dotted](root) to (leaf12);
    \draw[->](root) to (leaf13);
    \draw[->,decorate](leaf13) to (leaf22);
    \draw[->](leaf13.west) to (l3);
    \draw[->](leaf11.south) .. controls +(down:1cm) and +(left:1cm) .. (l3.west);
\end{tikzpicture}
\caption{Partial call graph that involves the critical functions to trigger the vulnerability. Another target in \texttt{xmalloc} is also included. The distance value of each function is annotated above the rectangular in red.}
\label{fig:motivation}
\vspace{-5pt}
\end{figure}
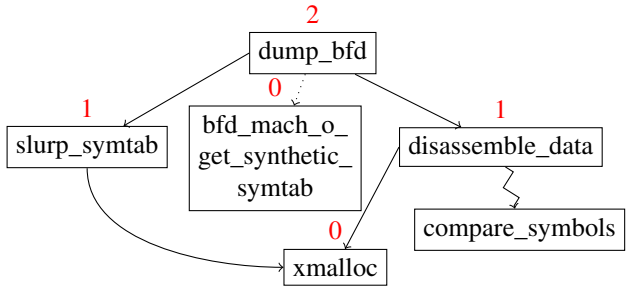

%% file: motivation_code.tex
\begin{listing}[!ht]
\begin{minted}[fontsize=\fontsize{7pt}{9pt}, xleftmargin=12pt,linenos]{c}
typedef struct bfd_symbol {
  struct bfd *the_bfd;
  /* other fields are omitted */
} asymbol;

asymbol *synthsyms;
long *(bfd_get_synthetic_symtab)(...);

void dump_bfd (...) {
  bfd_get_synthetic_symtab(..., &synthsyms);
  disassemble_data(...);
}

/* bfd_get_synthetic_symtab for Mach-O executable */
long bfd_mach_o_get_synthetic_symtab(..., asymbol **ret) {
  size_t size = count * sizeof(asymbol) + 1;
  char *s_start = bfd_malloc(size);
  *ret = (asymbol *) s_start;
  /* intialize each field of the asymbol structure,
     except for the the_bfd pointer field */  
}

/* disassemble_data finally calls compare_symbols */
int compare_symbols(const void *ap, const void *bp) {
  /* a and b point to the elements in synthsyms. */
  const asymbol *a = * (const asymbol **) ap;
  const asymbol *b = * (const asymbol **) bp;
  /* code to dereference a->the_bfd and b->the_bfd,
     resulting in access of uninitialized pointer */
}

/* Another target site detected by static analysis */
void * xmalloc (size_t size) {
  void *newmem = malloc (size);
  return (newmem);
}
\end{minted}
\caption{Simplified vulnerable code snippet.} 
\label{code:motivation}
\end{listing}

%% file: Overview.tex
In this section, we provide a brief design overview of \sysname{}, including its workflow (\cref{Workflow}) and the fuzzing loop (\cref{FuzzingLoop}).

\subsection{Workflow} \label{Workflow}

\begin{figure*}
\centering
    \centering
    \includegraphics[width=\textwidth]{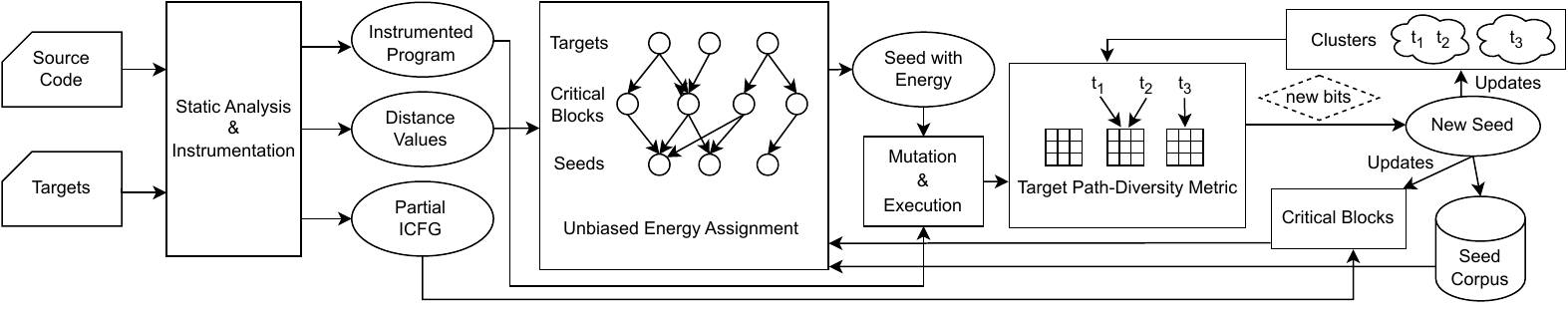}
    \captionsetup{skip=5pt}
    \caption{Workflow of \sysname{}.}
    \label{fig:overview}
\end{figure*}

Figure \ref{fig:overview} illustrates the workflow of \sysname{}.
In the static stage, \sysname{} first takes the program source code and target sites for static analysis and instrumentation. The static analysis generates a partial ICFG with vertices representing basic blocks that can reach at least one of the target blocks, along with distance values to these target blocks for each vertex. Each of these blocks is instrumented with our runtime callback function.
In the dynamic stage, our unbiased energy assignment generates energy values for all seeds, which determine the number of mutations and executions for each seed. For each execution of a mutated test case, we employ our target path-diversity metric to determine whether the test case should be stored in the corpus. If so, we also use it to update critical blocks and target clusters. Target clusters are utilized to group targets exhibiting similar behavior, thereby reducing the runtime overhead, particularly when the number of targets is substantial (discussed in \cref{TargetClustering}).

\subsection{Fuzzing Loop} \label{FuzzingLoop}

\begin{algorithm}[hbt!]
\footnotesize
\caption{Fuzzing Loop of \sysname{}} \label{FuzzingLoopAlgo}
\begin{algorithmic}[1]
\REQUIRE {Initial \emph{corpus}}
\REPEAT
    \STATE $seed\_energy$ := assign\_energy($E$, corpus)
    \FOR{$(seed, energy) \in seed\_energy$ in descending order}
        \FOR{$i$ from $1$ to $energy$}
            \STATE $s'$ := mutate\_input($seed$)
            \STATE $t'$ := execute($s'$)
            \IF{$t'$ crashes}
                \STATE add $s'$ to $crashes$
            \ELSIF{has\_new\_coverage($t'$)}
                \STATE add $s'$ to $corpus$
                \IF{$t'$ updates critical blocks}
                \STATE \textbf{break the outermost loop}
                \ENDIF
            \ENDIF
        \ENDFOR
    \ENDFOR
\UNTIL{timeout reached or abort-signal}
\RETURN $crashes$
\end{algorithmic}
\end{algorithm}

Algorithm \ref{FuzzingLoopAlgo} provides a high-level overview of the fuzzing loop in \sysname{}.
Each iteration of the outermost loop constitutes one cycle, during which all seeds in the corpus are supposed to be selected for fuzzing once.
At the beginning of each cycle, the \texttt{assign\_energy} function, our unbiased energy assignment, is invoked to assign an energy value to each seed. Unlike other greybox fuzzing approaches, including directed ones, our energy assignment algorithm operates globally for the entire corpus rather than locally for individual seeds. This global perspective is crucial for ensuring unbiased energy assignment.

In the second loop, we iterate through each seed-energy pair returned from \texttt{assign\_energy} in descending order based on energy, which allows \sysname{} to fuzz seeds with higher energy first. The seed is mutated using \texttt{mutate\_input}, the mutation algorithm from AFL++~\cite{Fioraldi2020AFLC}, and executed by the PUT with our instrumentation using \texttt{execute}.
The execution results $t'$ are examined similarly to other greybox fuzzing methods. However, \texttt{has\_new\_coverage} is now replaced by our target path-diversity metric, rather than the previous coverage metric used in AFL++.
Additionally, if a new seed updates the current critical blocks, the cycle is immediately halted to restart the next cycle. This is because our unbiased energy assignment depends on critical blocks; thus, if any updates occur with them, \sysname{} should discard the obsolete energy results and re-assign energy using the updated critical blocks.

%% file: Diversity.tex
In this section, we will discuss how our coverage metric improves the path diversity of already-covered targets. We introduce our multiple virgin maps in \cref{VirginMaps} and explain how we reduce overhead through clustering in \cref{TargetClustering}.

\subsection{Virgin Maps for Target Path Diversity} \label{VirginMaps}

\begin{algorithm}[hbt!]
\footnotesize
\caption{Coverage Metric for Test Case} \label{SaveIfInterestingAFLRun}
\begin{algorithmic}[1]
\REQUIRE {\emph{path\_trace} and \emph{virgin\_maps}}
\STATE store := \FALSE \\
\STATE data := set()
\FOR {i := 1 \TO N}
    \IF {path\_trace[i] $=$ 1}
        \STATE updates := set()
        \FOR {map $\in$ virgin\_maps}
            \IF {map[i] $=$ 1}
                \STATE map[i] := 0
                \STATE store := \TRUE
                \STATE updates.insert(map)
            \ENDIF
        \ENDFOR
        \IF {$|updates| \neq 0$}
            \STATE data.insert(updates)
        \ENDIF
    \ENDIF
\ENDFOR
\RETURN store, data
\end{algorithmic}
\end{algorithm}

Given a particular corpus state during a fuzzing campaign, the path diversity of a target is defined by the total coverage of seeds that cover this target. To enhance this path diversity, we propose to introduce an extra set of virgin maps.
As we discussed in \cref{CoverageGuidedFuzzing}, AFL++~\cite{Fioraldi2020AFLC} uses a virgin map to store the coverage information of the current seed corpus. When the execution trace of a test case covers any new bits not yet covered by the corpus, the virgin map is updated, and the test case is stored to the corpus. In our approach, rather than using a single virgin map, we maintain and compare the execution trace of each test case against multiple virgin maps.
These multiple virgin maps consist of the original virgin map used by AFL++, referred to as \emph{primary virgin map}, and a set of extra virgin maps derived from all targets covered by the test case, referred to as \emph{target virgin maps}.
After executing each test case, besides comparing its execution trace with primary virgin map like AFL++, we also compare it with our extra target virgin maps.
If any of these virgin maps indicate that new bits have been covered, the test case will be stored to the corpus, and the bits of corresponding virgin map will be marked as non-virgin.

This approach is detailed in Algorithm \ref{SaveIfInterestingAFLRun}.
The parameter $path\_trace$ is a bit array of length $N$, which is the execution trace of test case that has just been executed. The parameter $virgin\_maps$ is a set of virgin maps including primary virgin map and target virgin maps, each of which is also a bit array of length $N$. The algorithm is somewhat similar to the original one of AFL++, except we compare execution trace with multiple virgin maps instead of a single virgin map. Furthermore, for each bit location where an update occurs, we record all the virgin maps that have experienced an update at that bit location. While this information is not useful for determining whether to store the test case, it is useful for target clustering, which will be discussed later.

\subsection{Target Clustering} \label{TargetClustering}

Another question to consider is how to obtain the set of target virgin maps from a set of targets covered by a test case. A naive approach would be to maintain a target virgin map for each of the targets that have been covered by seeds in the corpus.
However, this approach could suffer from performance overhead: since comparing the execution trace with virgin maps is a frequent event, a large number of targets covered by each test case would require comparisons with a correspondingly large number of virgin maps, resulting in a non-negligible overhead.
Initially, we implemented this naive approach and, by observing how each target virgin map was updated with new bits from the execution trace, we discovered that some virgin maps frequently update the same bits simultaneously. The general idea is that we can cluster these virgin maps into a single virgin map. This can reduce the overhead caused by repeated comparisons while still maintaining the desired path diversity.

\vspace{2pt}\noindent\textbf{Obtaining clusters}. To select the set of clusters corresponding to the execution of a test case, \sysname{} merges all clusters containing the targets covered during the execution. Each cluster corresponds to one virgin map and vice versa. By assembling these clusters together with the primary cluster, we can get the set of virgin maps used in Algorithm~\ref{SaveIfInterestingAFLRun}. For an in-depth depiction, refer to Appendix \ref{ObtainingClusters}.

\vspace{2pt}\noindent\textbf{Association rule mining}.
Before discussing the details of our clustering algorithm, we will first introduce some concepts related to association rule mining used in this paper, which differ slightly from the original definitions. Assume we have a given set of items $I = \{i_1, i_2, ..., i_n\}$ and a database $D = \{t_1, t_2, ..., t_m\}$. Each element $t$ in $D$ is defined as a subset of $I$ (i.e., $\forall_{t \in D}\ t \subseteq I$). In other words, the database consists of multiple subsets of items.
In this paper, we define an association rule~\cite{Agrawal1993MiningAR} between two items, denoted as $i_a \Rightarrow i_b$, where $i_a \in I$ and $i_b \in I$. Intuitively, this rule suggests that when $i_a$ appears in a set $t$ within the database, $i_b$ is also likely to appear.
Several metrics have been defined to measure the extent to which a given rule holds in a given database. The support count of an itemset $X \subseteq I$ is the number of times the itemset appears in the database. In other words, the support count, denoted as $\sigma(X)$, is the number of $t \in D$ such that $X \subseteq t$. The confidence of a rule $i_a \Rightarrow i_b$ is a ratio measuring the frequency of $i_b$ appearing given that $i_a$ appears, calculated by $\frac{\sigma({\{i_a,i_b\}})}{\sigma({i_a})}$. In this paper, we consider a rule $i_a \Rightarrow i_b$ to be valid if both its support count $\sigma(\{i_a,i_b\})$ and its confidence are larger than a specified threshold.

When a new target is covered by the corpus, a new cluster containing only that target is created. At this point, the virgin map selection for this target is equivalent to the naive implementation mentioned earlier.
As the fuzzing campaign progresses, more dynamic information is gathered, enabling us to merge clusters together. Specifically, we use $data$ from all seeds in the corpus, collected through Algorithm \ref{SaveIfInterestingAFLRun}, as a database for association rule mining~\cite{Agrawal1993MiningAR}. This allows \sysname{} to identify target virgin maps that can be clustered together.
In Algorithm \ref{SaveIfInterestingAFLRun}, we can observe that $data$ consists of sets of clusters, as each virgin map corresponds to a cluster. We treat each cluster as an item and consider the union of $data$ from all seeds in the corpus as the database in the context of association rule mining. Additionally, since our dataset is constructed by joining the dataset of each individual seed, we denote the dataset of a single seed $s$, which is a subset of the entire dataset, as $I_s$.

Algorithm \ref{UpdateSupportCount} outlines the process of updating support count after a new seed is added to the corpus. The parameter $data$ comes from the return value of Algorithm \ref{SaveIfInterestingAFLRun}, and the parameter $support\_count$ is a global map used to record the support count of each itemset with one or two elements. For each set in $data$, we count both single items and pairs of items by $\frac{1}{|data|}$ instead of $1$. This design choice is made to prevent bias caused by some seeds that trigger many new bits. (Remember that each element in $data$ corresponds to a new bit.) As a result, we only assign a total support count of $1$ for each seed, which is then distributed uniformly among all sets of items it contains.

\begin{algorithm}[hbt!]
\footnotesize
\caption{Support Count Update for Each New Seed} \label{UpdateSupportCount}
\begin{algorithmic}[1]
\REQUIRE {\emph{data} and \emph{support\_count}}

\FOR {items $\in$ data}
    \FOR {each single item $i$ $\in$ items}
        \STATE support\_count[$\{i\}$] += $\frac{1}{|data|}$
    \ENDFOR
    \FOR {each pair of items $\{i_a,i_b\}$ $\subseteq$ items}
        \STATE support\_count[$\{i_a, i_b\}$] += $\frac{1}{|data|}$
    \ENDFOR
\ENDFOR

\end{algorithmic}
\end{algorithm}

Using the support count from Algorithm \ref{UpdateSupportCount}, we can calculate the confidence value of a given rule $i_a \Rightarrow i_b$ using the formula described earlier.

\vspace{2pt}\noindent\textbf{Clustering}. Given an association rule $i_a \Rightarrow i_b$, if both its support count and confidence are higher than the threshold values, we can consider it to be valid. This semantically implies that if the execution trace of a new seed covers a new virgin bit in $i_a$, it will typically also cover the same new virgin bit in $i_b$. In other words, the behavior of virgin bit updates in $i_a$ is very similar to that of $i_b$, so we can merge such redundant cluster $i_a$ into $i_b$ to reduce overhead while keeping the capability of improving path diversity of covered targets. Finally, it is important to note that the primary cluster cannot be merged into any other cluster, but other clusters can be merged into it.

%% file: Weighted.tex
In this section, we describe the definition of critical blocks for each target (\cref{CriticalCoverage}) and the process of selecting favored seeds (\cref{SeedSelection}).
We then introduce how \sysname{} assigns energy among seeds at the beginning of each cycle with such information (\cref{EnergyAssignment}).

\subsection{Critical Blocks} \label{CriticalCoverage}

We firstly define a set $R$ as the set of all basic blocks that have been covered by the current seed corpus, with $v \in R$ denoting a covered block within the set. Next, we define \emph{critical blocks} $C_t$ for each target $t$ as a subset of $R$.
The underlying idea is to identify a set of blocks for each target at a given state, so that we should concentrate our fuzzing energy on seeds that cover these blocks, ultimately enabling more efficient exploration or exploitation of the target.
In this section, we introduce this concept.

\vspace{2pt}\noindent\textbf{For covered targets}. During a fuzzing campaign, we can categorize the set of all targets into two groups: one set consists of targets already covered by the current corpus (i.e., $t \in R$), and the other set contains targets not yet covered by the current corpus (i.e., $t \notin R$). If a target is already covered, the critical block for this target is simply the target block itself. Consequently, to fuzz this target, we should focus on fuzzing seeds that cover such a target.

\vspace{2pt}\noindent\textbf{For uncovered targets}.
If a target is not yet covered, we should instead attempt to find a seed that covers the target by fuzzing seeds in the current corpus. Drawing on the intuition from \cite{She2022EffectiveSS},
we define the critical blocks for such a target as the blocks that lie at the \emph{boundary between the explored parts and unexplored parts of the path toward the target} in the ICFG, known as critical boundary blocks.

We firstly need to define some notations to introduce the exact definition of critical blocks. Given a target basic block $t$, we can define a path\footnote{Here, the term ``path" does not refer to the program's execution path. Instead, it refers to the concept of a path as defined in graph theory.} toward the target in the ICFG as follows:

$$v_1 \rightarrow v_2 \rightarrow ... \rightarrow v_{n} \rightarrow t$$

In the path above, $v_i$ with $i \in \{1, ..., n\}$ are basic block vertices in the ICFG, and each arrow is either a control flow edge or a call edge that connects two adjacent vertices in the ICFG. The idea behind critical boundary blocks is to find each block visited by the corpus such that there exists a path from this block to target, and all vertices along this path are basic blocks not yet visited by the corpus except for the block itself. This definition aligns with the intuition of finding covered blocks at the boundary mentioned earlier, with the uncovered blocks considered as the unexplored part. Formally, critical blocks of an uncovered target $t$ can be defined as all blocks $v_1 \in R$ such that there exists a path from $v_1$ to $t$ with $\forall_{v \in \{v_2, v_3, ..., v_n, t\}} v \notin R$.

\vspace{2pt}\noindent\textbf{Example}. We provide an example to illustrate concept of critical blocks. Figure \ref{ExampleGraph} is the ICFG of a program, with $1$ and $2$ as target blocks. We now have three seeds in corpus with paths $A \rightarrow B \rightarrow C \rightarrow H$, $A \rightarrow B \rightarrow D \rightarrow E$ and $A \rightarrow K \rightarrow 2$. We mark basic blocks covered by the corpus in gray. Among these blocks, $C$ and $D$ are critical blocks of target $1$, because we can have two paths to target $C \rightarrow 1$ and $D \rightarrow F \rightarrow G \rightarrow 1$ respectively, with $F$, $G$, $1$ not yet covered by the corpus. Furthermore, $2$ itself is the critical block of target $2$, because the target has already been covered by the corpus. %

\input{flow_figure.tex}

\subsection{Favored Seed Selection} \label{SeedSelection}

Besides a target virgin map for each cluster introduced in \cref{TargetClustering}, \sysname{} also maintains a top-rated array for each cluster, similar to the original top-rated array discussed in \cref{CoverageGuidedFuzzing}. When a new seed is added to the corpus, a set of top-rated arrays are selected using Algorithm \ref{SelectVirginMaps}, and we update these top-rated arrays using the same method as AFL++.

Using these top-rated arrays, we can select a set of favored seeds.
We can apply the queue culling algorithm, similar to AFL++, on top-rated arrays of all clusters to select the favored seeds. For each top-rated array, we use the queue culling algorithm from AFL++ to select a set of favored seeds, and we then take the union of each set of favored seeds from each top-rated array to obtain the total set of favored seeds.

\subsection{Energy Assignment} \label{EnergyAssignment}

In this section, we will introduce the method for assigning energy among seeds using critical blocks of each target.
In previous directed fuzzing works, a distance-to-target metric for each seed is used as the oracle for energy assignment. By contrast, instead of using such distance oracle, we utilize \emph{critical blocks covered by each seed} as the oracle for energy assignment.
The primary concept behind our algorithm involves distributing uniform weight from targets through critical blocks to seeds.
Our goal is to assign energy to each seed at the beginning of a cycle so that the total energy of each seed after assignment can approach the proportion of such seed weight as much as possible.

\vspace{2pt}\noindent\textbf{Target weight}. As mentioned earlier, \sysname{} starts with uniform weight values for all target basic block. Specifically, we have a vector of ones representing the weight for each target, denoted as $\mathbf{1}$. However, our approach does allow users to specify a customized weight value for each target. We provide a detailed description of this by-product in Appendix \ref{TargetWeightCalculation}.

\vspace{2pt}\noindent\textbf{Block weight}.
In this part, we discuss how we distribute weight from each target to its critical blocks.
For each target $t$, we have a set of critical blocks $C_t \subseteq R$.
By taking the union of $C_t$ for all targets, we obtain the set of all critical blocks $C \subseteq R$, where $R$ is the set of all covered blocks defined earlier. We can now define a matrix $B$ for weight distribution:
each column of $B$ represents a corresponding target, and each row of $B$ represents a block in $C$.
An element of the matrix at row $b$ and column $t$ is defined as follows:

$$
B_{b, t} =
\begin{cases}
 \frac{1}{distance[b, t] + k} & \text{if $b \in C_t$} \\
 0 & \text{otherwise}
\end{cases}
$$

In the equation, the term $distance[b, t]$ represents the distance from the basic block $b$ to the target $t$, computed using the method described in Appendix \ref{DistanceComputation}, because this distance computation is only an auxiliary component of our approach.
The constant $k$ is a positive value that controls how significantly the distance value influences the preference for a block. The larger the value of $k$, the smaller the effect will be.
Next, the matrix $B$ is normalized to $\hat{B}$, ensuring that each column of the matrix sums to $1$.
Now, the distribution of the set of critical blocks can be obtained by calculating the product $\hat{B}\mathbf{1}$, which distributes weight from each target to each critical block.

\vspace{2pt}\noindent\textbf{Seed weight}. The process of distributing weight from each critical block to seeds is quite similar to the distribution mentioned earlier. We also have a matrix $S$ for this distribution: each column of $S$ represents a critical block in $C$, while each row of $S$ represents a seed in the corpus. An element at row $s$ and column $b$ can be defined as follows:

$$
S_{s, b} =
\begin{cases}
 score[s] & \text{if favored $s$ covers $b$} \\
 0.05 \cdot score[s] & \text{if unfavored $s$ covers $b$} \\
 0 & \text{otherwise}
\end{cases}
$$

Next, matrix $S$ is also normalized to $\hat{S}$ to ensure that the sum of each column is equal to $1$. The term $score[s]$ represents the energy assigned to seed $s$ when fuzzed in coverage-guided mode, as computed by the algorithm from AFL++.
This energy can, in some sense, represent the preference of a seed from AFL++'s perspective. We utilize this preference when distributing weight from each critical block to seeds.
Additionally, AFL++ skips unfavored seeds with a $95\%$ probability. We also follow this design by calculating an expected energy for unfavored seeds. The favorability of a seed is determined using the favored seeds selected. Finally, we can calculate the seed weight vector by computing $\hat{S}\hat{B}\mathbf{1} + c$. This final seed weight vector is also normalized to $r$, representing the final ratio to be approached.

\vspace{2pt}\noindent\textbf{Preventing local optima}. One problem of our energy assignment based on critical blocks is the potential for local optima. To mitigate such issue, during the calculation of the seed weight vector, we add a vector $c$, which represents the energy assignment of each seed in coverage-guided fuzzing. The sum of $c$ is calculated by multiplying the total energy (i.e., the number of targets) by a small fraction.

\vspace{2pt}\noindent\textbf{Seed energy assignment}. Now, \sysname{} can assign energy to seeds using the obtained $r$. \sysname{} also maintains a vector $b$ representing the energy that has been assigned to each seed previously. Given a total energy value $E$ for this cycle, we assign energy $E$ to each seed in the form of vector $x$, such that $x + b$ can approach the ratio $r$ as closely as possible, under the constraint $\sum_i x_i = E$. Instead of using a general algorithm of mathematical optimization, we develop an algorithm with a worst-case complexity of $O(n^2)$ that guarantees the optimal solution.
We discuss this algorithm in Appendix \ref{MathematicalOptimizationAlgorithm}, as it is unrelated to fuzzing.

%% file: flow_figure.tex
\begin{figure}
\centering
\scalebox{0.575}
{
\tikzstyle{leaf}=[circle,draw=black]
\tikzstyle{arrow}=[->,thick]
\begin{tikzpicture}
    \node[leaf,fill=gray!50](00){A};
    \node[leaf,below of=00,xshift=-1.6cm,fill=gray!50](10){B};
    \node[leaf,below of=00,xshift=1.6cm,fill=gray!50](11){K};
    \node[leaf,below of=10,xshift=-0.8cm,fill=gray!50,ultra thick](20){\textbf{C}};
    \node[leaf,below of=10,xshift=0.8cm,fill=gray!50,ultra thick](21){\textbf{D}};
    \node[leaf,below of=11,xshift=-0.8cm,fill=gray!50,ultra thick](22){\textbf{2}};
    \node[leaf,below of=11,xshift=0.8cm](23){L};     
    \node[leaf,below of=20,xshift=-0.4cm,fill=gray!50](31){H};
    \node[leaf,below of=21,xshift=-0.4cm](32){F};
    \node[leaf,below of=21,xshift=0.4cm,fill=gray!50](33){E}; 
    \node[leaf,below of=32,xshift=-0.6cm](41){G};
    \node[leaf,below of=32,xshift=0.6cm](42){I};
    \node[leaf,below of=41,xshift=-0.4cm](50){1};
    \node[leaf,below of=41,xshift=0.4cm](51){J};
    \draw[arrow](00)--(10);
    \draw[arrow](00)--(11);
    \draw[arrow](10)--(20);
    \draw[arrow](10)--(21);
    \draw[arrow](11)--(22);
    \draw[arrow](11)--(23);
    \draw[arrow](20)--(50);
    \draw[arrow](20)--(31);
    \draw[arrow](21)--(32);
    \draw[arrow](21)--(33);
    \draw[arrow](32)--(41);
    \draw[arrow](32)--(42);
    \draw[arrow](41)--(50);
    \draw[arrow](41)--(51);
\end{tikzpicture}
}
\caption{Inter-procedural Control Flow Graph} \label{ExampleGraph}
\end{figure}
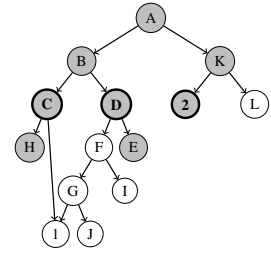

%% file: Evaluation.tex
In this section, we explore the following research questions through experiments.
\textbf{RQ1}: How effective is \sysname{} at triggering vulnerabilities compared to other fuzzers and ablated versions of \sysname{}? (\cref{Comparison} and Appendix \ref{AdditionalExperiments})
\textbf{RQ2}: How effective is \sysname{} in ensuring fairness for each target? (\cref{TargetFairness})
\textbf{RQ3}: What is the runtime overhead incurred by \sysname{} during fuzzing campaign? (\cref{Overhead})
\textbf{RQ4}: Can \sysname{} discover new zero-day vulnerabilities in programs extensively fuzzed by OSS-Fuzz using state-of-the-art industrial fuzzers? (\cref{ZeroDayVuln})
\textbf{RQ5}: Do the extra seeds generated by \sysname{} assist in triggering vulnerabilities? (Appendix \ref{QualityExtra})

\vspace{2pt}\noindent\textbf{Experiment setup}. Unless otherwise specified, all experiments were conducted on an Ubuntu 20.04.5 LTS machine equipped with an Intel(R) Xeon(R) Gold 5318S CPU featuring 96 logical cores and 256 GB of RAM.

\subsection{Fuzzer Comparison} \label{Comparison}

We use two sets of vulnerabilities for experiments. Firstly, to compare \sysname{} with other fuzzers, we employ the Magma~\cite{Hazimeh2020MagmaAG}, a benchmark commonly used by many fuzzing works, to evaluate multi-target directed fuzzing for bug reproduction. Additionally, to demonstrate the individual contribution from each component of \sysname{}, we use another set of vulnerabilities in a real-world bug discovery scenario to conduct an ablation study.

\smallskip\noindent\textbf{Settings}. We set the confidence threshold to $100\%$ to only cluster target blocks that consistently dominate each other, because the target sites do not cause significant overhead in these experiments.
We also individually remove two key components from \sysname{}, resulting in variations denoted as \sysname{}- and \sysname{}*.
\sysname{}- refers to \sysname{} devoid of the target path-diversity metric, while \sysname{}* denotes \sysname{} employing AFL++'s energy assignment rather than our unbiased energy assignment.
Following the fuzzing papers' general experimental approach, we ran each fuzzing campaign for 24 hours and repeated %
10 times. To prevent interference between experiments, we left some CPU cores unused, in line with previous research practices.

\subsubsection{Evaluation on Magma} \label{MagmaEvaluation}

Magma operates by reintroducing multiple 1-day bugs into real-world programs using patches, enabling us to set multiple targets based on these patches and simultaneously fuzz multiple bugs. We excluded some bugs that can be found in less than ten minutes by coverage-guided fuzzers, as these bugs can also be discovered by directed fuzzers while still in coverage-guided mode, which would not effectively demonstrate the efficacy of directed fuzzing.

\vspace{2pt}\noindent\textbf{Fuzzers for comparison}.
We compare \sysname{} with seven fuzzers: AFL++~\cite{Fioraldi2020AFLC}, AFLGo~\cite{Bhme2017DirectedGF}, Hawkeye~\cite{Chen2018HawkeyeTA}, Parmesan~\cite{sterlund2020ParmeSanSG}, FishFuzz~\cite{Zheng2022FishFuzzTL}, WindRanger~\cite{Du2022WindrangerAD} and MOpt~\cite{Lyu2019MOP}. We chose AFL++ because it is the grey box fuzzer upon which we build \sysname{}, and selected AFLGo and Parmesan as they are open-source directed fuzzers commonly used for comparison. MOpt was chosen due to its outstanding performance on the Magma benchmark, where it currently leads as the most effective fuzzer. We also included Hawkeye, FishFuzz and WindRanger, as they all attempt to address the global optimum discrepancy problem. However, since Hawkeye is not open-source, we implemented our own prototype based on the approaches described in the paper. We exclude works that are orthogonal to our approach~\cite{Srivastava2022OneFD,luo2022selectfuzz,Huang2022BEACONDG,Zong2020FuzzGuardFO}, because these works can be applied on top of \sysname{} without modifying both methods.

\vspace{2pt}\noindent\textbf{More fine-grained ablation study}.
In addition to \sysname{}-, we introduced a more fine-grained ablation variant named \sysname{}-{}-. This ablated version not only removes the target path-diversity metric but also discards the concept of critical block. Instead of applying critical blocks, \sysname{}-{}- assigns energy using all previously covered blocks that can reach the target via the ICFG. The purpose of this ablation study is to underscore the impact of the critical block design.

\vspace{2pt}\noindent\textbf{Target locations}. Magma inserts one or more \texttt{MAGMA\_LOG} macros into the tested program for each bug to notify when the bug is covered or triggered during execution. We set the locations of these macros as targets for each bug. Furthermore, Magma patches the program to invert the commits that fix the bugs, and we set these locations as targets as well.

\input{eval_table.tex}

\vspace{2pt}\noindent\textbf{Results}. The results are presented in Table \ref{MagmaBenchmarkResults}. Each entry in the table represents the average exposure time required for the fuzzer to trigger the vulnerability, calculated using survival analysis. We also computed p-values using the Mann-Whitney U test~\cite{Mann1947OnAT}, employing the alternative hypothesis that \sysname{} has a shorter exposure time. P-values are included in the parenthesis. T.O. indicates that the bug was not triggered in any fuzzing campaign. Note that Parmesan and WindRanger failed to run on some programs at either compile time or runtime, even after we fixed some of the bugs, so we mark those entries as N/A. We observe that \sysname{}(-) outperforms other counterparts in most cases. On average, using the geometric mean, \sysname{} achieves speedup factors of $168\%$, $109\%$, $235\%$, $183\%$, $147\%$, $157\%$, $78\%$ and $56\%$, and discovers $38\%$, $20\%$, $138\%$, $29\%$, $24\%$, $14\%$, $50\%$ and $16\%$ more vulnerabilities compared to AFL++, AFLGo, Parmesan, FishFuzz, Hawkeye, WindRanger, MOpt and \sysname{}-{}-, respectively. Besides, according to the results, \sysname{} can discover a bug that other fuzzers so far have not triggered, LUA002.
We observe that the effectiveness of \sysname{}-{}- diminished with the ablation; however, it still outperforms other fuzzers.
This demonstrates that the design of the critical block indeed contributes to its enhanced effectiveness.
Moreover, the fact that \sysname{}-{}- still surpasses other fuzzers highlights the value of its unbiased energy assignment, even in the absence of the critical block design.

\subsubsection{Evaluation on Real Scenario}

\input{ablation_table.tex}

\begin{figure}
\centering
    \centering
    \includegraphics[scale=0.2]{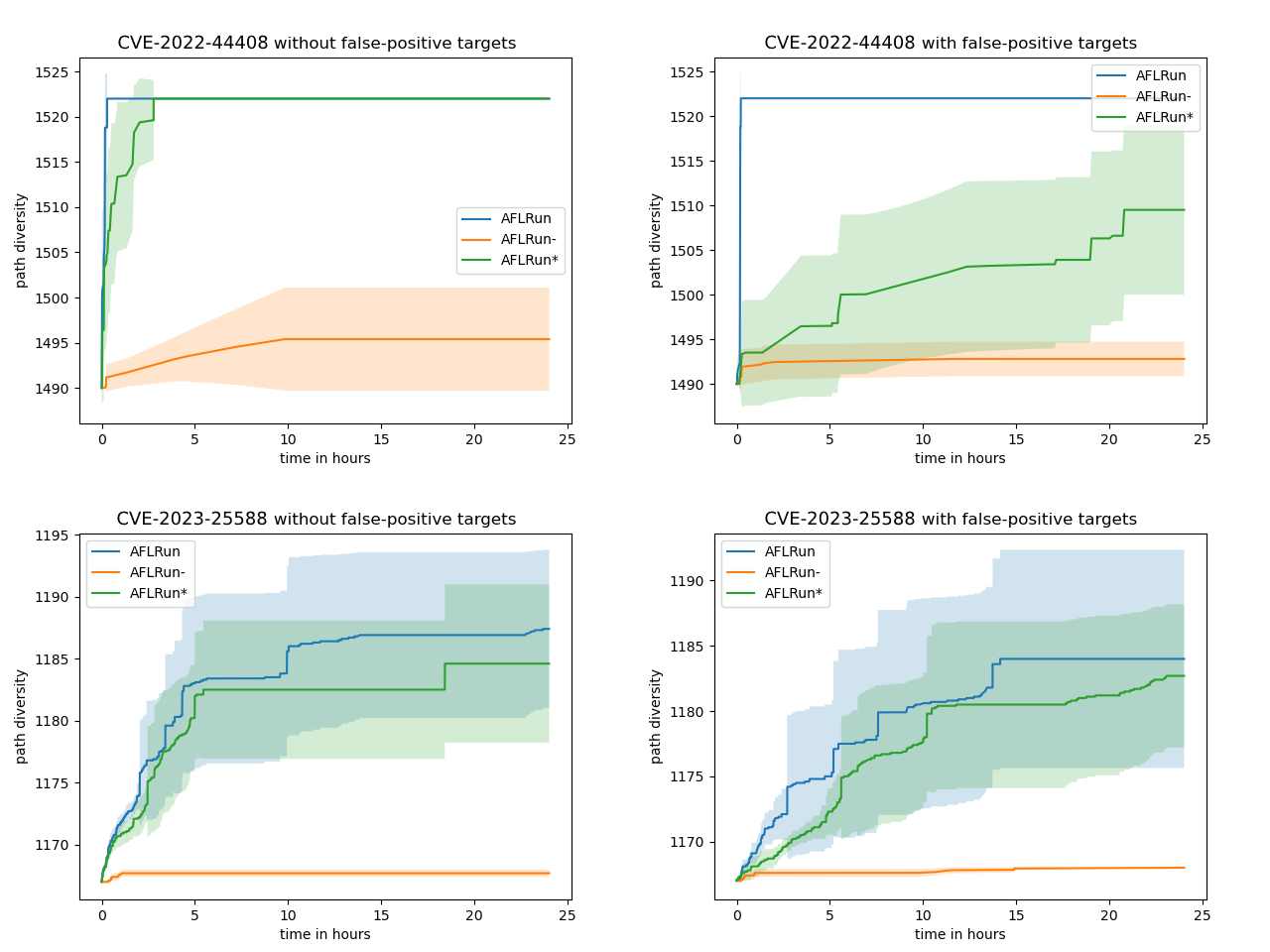}
    \caption{Path diversity curves}
    \label{PathDiversityCurve}
    \vspace{-5pt}
\end{figure}

The results from the Magma benchmark reveal that \sysname{} does not significantly outperform \sysname{}-.
Upon further investigation, we discovered that most bugs in Magma are quite simple, probably because Magma requires easy detection of bug trigger. However, our target path-diversity metric is designed to handle complex bugs that require path diversity for triggering, thus the benchmark results fail to fully illustrate the impact of our method.
Moreover, the Magma experiments are designed to test bug reproduction capability when the bug sites are known. Yet, in real vulnerability discovery scenario
such as the one detailed in \cref{ExampleVuln}, numerous false positive targets might also exist. This situation has not been evaluated in the aforementioned benchmarking experiment.

To address this deficiency of Magma benchmark, we conduct an additional 
experiment in a different setting, using some bugs that align with our scenario. 
To demonstrate the impact of encountering false-positive targets (emulating real-world bug detection), we carry out the experiment under two conditions: with and without the introduction of false-positive targets. 
Those false-positive targets are
randomly chosen from the bug reports detected by Clang static analyzer~\cite{ClangChecker}. 

We did not include \sysname{}-{}- in this experiment as its primary goal was to demonstrate \sysname{}'s effectiveness on complex bugs that require path diversity for activation. In contrast, \sysname{}-{}- was designed to demonstrate the contributions of different components of unbiased energy assignment in the Magma experiment.

The bug triggering time results are displayed in Table \ref{AblationResults}. We observe that \sysname{} significantly outperforms \sysname{}-. Furthermore, with the introduction of false-positive targets, \sysname{} generally exhibits superior bug-triggering capability compared to \sysname{}*. This indicates that the unbiased energy assignment indeed helps in managing the seed explosion caused by the increased target count.

These results are further elucidated by graphing the path diversity curves from several experiments. For the corpus created by each fuzzing campaign, we track the increase in path diversity of basic block coverage necessary to trigger the vulnerability over time. For each set of repeated experiments, we compute the mean and the corresponding confidence interval. The outcomes are depicted in Figure \ref{PathDiversityCurve}. As seen in the figure, the target path-diversity metric significantly enhances the path diversity of the target.

\subsection{Target Fairness Comparison} \label{TargetFairness}

In this experiment, we evaluate the effectiveness of \sysname{} in balancing each target compared to its counterparts, addressing RQ2.
We aim to scale up the total number of targets by utilizing recently introduced commits as target sites for each program in the Magma benchmark~\cite{Hazimeh2020MagmaAG}.
We also disable the bugs introduced by Magma, as we are not concerned with their triggering in this experiment. Furthermore, since the primary goal of this experiment is to assess the effectiveness of our unbiased energy assignment, we have ablated the target path-diversity metric (i.e., we use \sysname{}-).

\begin{figure}
\centering
    \centering
    \includegraphics[scale=0.25]{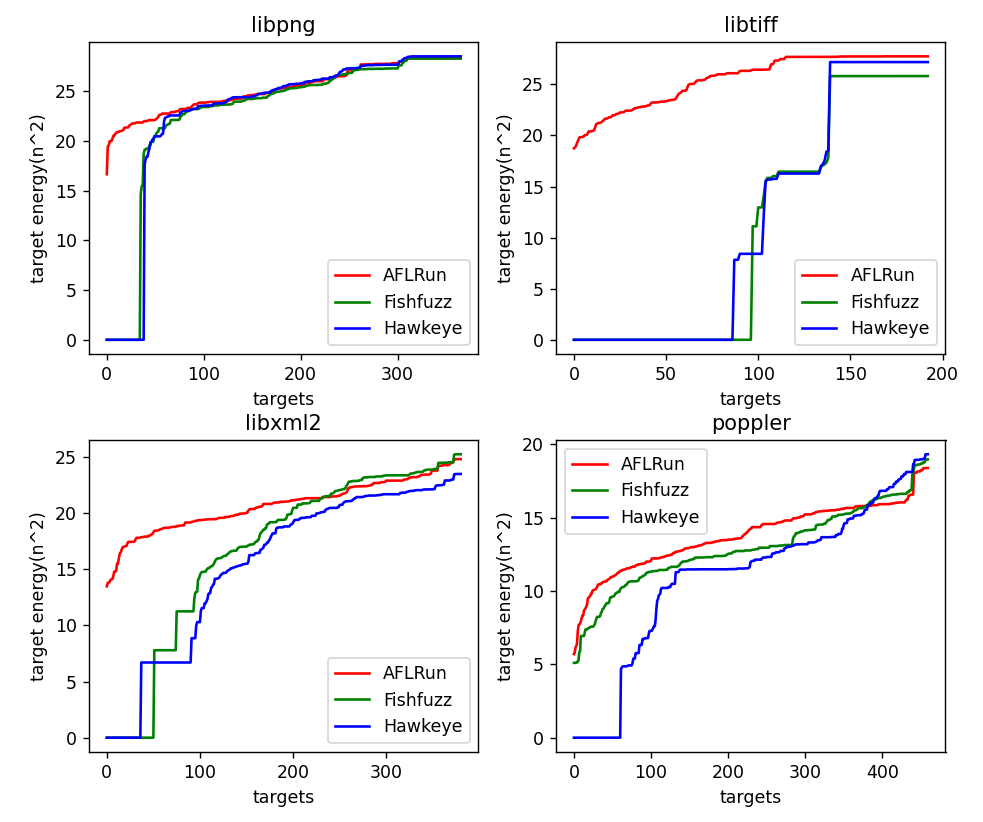}
    \caption{Energy spent on each target}
    \label{fig:evaluation}
    \vspace{-5pt}   
\end{figure}

We conducted the experiment for three directed fuzzers: Hawkeye, FishFuzz, and \sysname{}-. For each fuzzer, we carried out an 8-hour fuzzing campaign, repeating each experiment 10 times.
We recorded the energy assigned (i.e., the number of mutations and executions performed) to each seed. Additionally, we used a modified version of \texttt{afl-showmap} to determine the target coverage of each seed. With this information, we computed the total energy spent on fuzzing each target. We sorted the energy for each target and plotted a graph similar to the one in FishFuzz~\cite{Zheng2022FishFuzzTL}. The results are illustrated in Figure \ref{fig:evaluation}.
The x-axis represents all targets ordered incrementally by energy spent on them, and the y-axis is the corresponding energy.
We can see that \sysname{} generates flatter curves compared to other counterparts, indicating a more balanced energy assignment for each target, especially for \texttt{libtiff} and \texttt{libxml2}.

\subsection{Overhead Measurement} \label{Overhead}

To evaluate the overhead incurred by \sysname{} at runtime in response to RQ3, we also scale up the total number of targets by utilizing target sites of recent commits from \cref{TargetFairness}. Similarly, the fuzzing campaign is conducted with the vulnerabilities introduced into these programs disabled.

To demonstrate the effectiveness of our clustering algorithm, we conducted overhead experiments under five different configurations: for three of them, we enabled target clustering with varying confidence threshold values of $50\%$, $70\%$, and $90\%$ and a consistent support count threshold value of $500$; for one of them, the clustering was disabled, meaning each target corresponds to a distinct virgin map and top-rated array; for the last one, we removed our path diversity component (\sysname{}-). We ran each experiment for 3 hours, which was repeated 20 times. We recorded the number of clusters at the end of each fuzzing campaign and calculated the arithmetic mean across all repeated experiments for each configuration. Similarly, we profiled the proportion of time spent in \sysname{} process to total time for each fuzzing campaign and calculated the mean. The results are shown in Table \ref{OverheadResults}. Additionally, we measured the overhead incurred by the instrumented code. However, such overhead is negligible: it occupies less than $0.01\%$ of total time in every fuzzing campaign.

\begin{table}[]
\centering
\caption{Overhead Results: The upper values of each program are average number of clusters; the lower values are time ratio spent in the PUT.}
\label{OverheadResults}
\resizebox{!}{2.0cm}{
\begin{tblr}{
  cells = {c},
  cell{2}{1} = {r=2}{},
  cell{4}{1} = {r=2}{},
  cell{6}{1} = {r=2}{},
  cell{8}{1} = {r=2}{},
  cell{10}{1} = {r=2}{},
  cell{12}{1} = {r=2}{},
  vlines,
  hline{1-2,4,6,8,10,12,14} = {-}{},
  hline{3,5,7,9,11,13} = {2-6}{},
}
Program    & Threshold = 50\% & Threshold = 70\% & Threshold = 90\% & w/o Cluster  & w/o Diversity  \\
libpng & 157.85 & 159.35 & 157.05 & 327.05 & 1.0 \\
 & 86.03\% & 86.07\% & 86.31\% & 80.41\% & 93.26\% \\
libsndfile & 303.05 & 302.35 & 307.95 & 332.6 & 1.0 \\
 & 97.60\% & 97.66\% & 97.62\% & 97.08\% & 98.33\% \\
libtiff & 188.75 & 190.2 & 189.25 & 188.3 & 1.0 \\
 & 78.27\% & 80.00\% & 77.29\% & 81.97\% & 91.24\% \\
libxml2 & 156.45 & 171.95 & 188.25 & 366.15 & 1.0 \\
 & 96.85\% & 96.97\% & 96.79\% & 95.46\% & 98.23\% \\
lua & 342.2 & 326.85 & 339.2 & 335.2 & 1.0 \\
 & 97.73\% & 97.69\% & 97.70\% & 97.75\% & 99.45\% \\
poppler & 159.95 & 169.35 & 164.85 & 173.65 & 1.0 \\
 & 99.06\% & 99.08\% & 99.07\% & 98.98\% & 99.48\%
\end{tblr}
}
\end{table}

Each row in the table represents a different configuration. The first three columns represent experiments with varying confidence threshold values; the fourth column represents an experiment without the clustering algorithm; and the last column represents an experiment with \sysname{}-. Each entry displays a pair of values: the upper value represents the average number of clusters when the fuzzing campaigns terminate, and the lower value is the average ratio of time spent executing the PUT over the total time of each fuzzing campaign. We can observe that the overhead incurred by the \sysname{} process is not significant compared to the time spent in executing the PUT. Additionally, the clustering algorithm that reduces the total number of virgin maps significantly does decrease some overhead spent in the fuzzer process.

\subsection{Vulnerability Discovery} \label{ZeroDayVuln}

Lastly, we present all vulnerabilities found by \sysname{} to answer RQ4. We use \sysname{} to fuzz the latest versions of some famous programs that have been continuously fuzzed by the OSS-Fuzz project~\cite{OSSFuzz} using powerful Google machines. We integrate \sysname{} into OSS-Fuzz to simplify fuzzing each program without struggling with the environment. Since \sysname{} is a directed fuzzer, we set target locations using two approaches. First, we attempt to fuzz recently introduced commits by setting recently changed code locations as targets. The intuition behind this is that recently modified code may be more error-prone because older code has already been thoroughly fuzzed by OSS-Fuzz, while new code is less fuzzed. Second, we use the static analysis tool to identify suspicious code locations. Specifically, we write CodeQL~\cite{Moor2007KeynoteA} queries to find potentially vulnerable code locations as our targets.

\begin{table}[h!]
\caption{Zero-day vulnerabilities found by \sysname{}.}
\label{NewBugs}
\resizebox{\columnwidth}{!}{%
\begin{tabular}{cccc}
\hline
Project & ID & Type & Status \\ \hline
wabt & CVE-2022-44407 & Out-of-bounds Read & Fixed \\
wabt & CVE-2022-44408 & Type Confusion & Fixed \\
wabt & CVE-2022-44409 & Out-of-bounds Read & Fixed \\
skia & issue-40045089 & Uncontrolled Recursion & Accepted \\
freetype & issue-1159 & Uncontrolled Recursion & Fixed \\
binutils & CVE-2023-25585 & Access of Uninitialized Pointer & Fixed \\
binutils & CVE-2023-25587 & NULL Pointer Dereference & Fixed \\
binutils & CVE-2023-25588 & Access of Uninitialized Pointer & Fixed \\
binutils & CVE-2023-25586 & Use of Uninitialized Variable & Fixed \\
binutils & CVE-2023-25584\tablefootnote{This CVE consists of a group of 20 vulnerabilities in total. We conduct a case study for this CVE and list each of the individual vulnerability IDs in the Appendix \ref{CVE202384}.} & Out-of-bounds Read & Fixed \\ \hline
\end{tabular}%
}
\end{table}

We discovered 29 previously unknown vulnerabilities in these programs, which are listed in Table \ref{NewBugs}. We responsibly reported them to the vendors. All vulnerabilities marked as Fixed have already been addressed by the developers in the upstream. At the time of writing, 28 vulnerabilities have been fixed, and 8 CVE IDs have been assigned.

%% file: eval_table.tex
\begin{table*}[h]
\caption{Magma Benchmark Results}
\label{MagmaBenchmarkResults}
\footnotesize
\centering
\setlength\tabcolsep{3pt}
\begin{tabular}{ccccccccccc}
\toprule
Bug ID & AFL++ & AFLGo & Parmesan & FishFuzz & Hawkeye & WindRanger & MOpt & \sysname{}-{}- & \sysname{} & \sysname{}- \\
\midrule

LUA002 & T.O.(0.18) & T.O.(0.18) & T.O.(0.18) & T.O.(0.18) & T.O.(0.18) & N/A & T.O.(0.18) & T.O.(0.18) & T.O. & \textbf{23.32h} \\ %
LUA004 & 4.23h(1.00) & 8.69h(0.96) & T.O.(<.01) & 5.93h(0.99) & \textbf{3.37h}(1.00) & N/A & 7.47h(0.97) & 13.95h(0.61) & 15.41h & 20.42h \\ %
PDF002 & T.O.(0.04) & T.O.(0.04) & N/A & T.O.(0.04) & 21.72h(0.14) & 21.32h(0.23) & 22.26h(0.12) & 20.90h(0.35) & \textbf{17.10h} & 21.65h \\ %
PDF003 & 6.60h(<.01) & 3.84h(0.03) & N/A & 5.74h(0.01) & 7.78h(<.01) & 5.58h(0.02) & 7.48h(<.01) & \textbf{88.99m}(0.95) & 4.04h & 2.15h \\ %
PDF006 & \textbf{17.49h}(0.95) & T.O.(0.18) & N/A & T.O.(0.18) & T.O.(0.18) & T.O.(0.18) & T.O.(0.18) & 20.59h(0.79) & 22.81h & T.O. \\ %
PDF011 & 22.83h(0.86) & T.O.(1.00) & N/A & T.O.(1.00) & T.O.(1.00) & 22.74h(0.86) & \textbf{21.14h}(0.94) & T.O.(1.00) & T.O. & T.O. \\ %
PDF014 & T.O.(0.08) & T.O.(0.08) & N/A & T.O.(0.08) & T.O.(0.08) & 22.67h(0.29) & T.O.(0.08) & T.O.(0.08) & \textbf{21.32h} & 21.69h \\ %
PDF018 & 14.87h(<.01) & 21.67h(<.01) & N/A & 19.37h(<.01) & 22.48h(<.01) & 2.59h(<.01) & 40.01m(0.30) & 2.79h(<.01) & \textbf{36.39m} & 3.13h \\ %
PDF019 & 20.51h(0.22) & 21.63h(0.06) & N/A & T.O.(<.01) & T.O.(<.01) & 23.64h(<.01) & 21.26h(0.08) & 19.27h(0.28) & T.O. & \textbf{18.44h} \\ %
PDF021 & 23.89h(0.50) & 23.47h(0.50) & N/A & T.O.(0.18) & T.O.(0.18) & 22.73h(0.50) & T.O.(0.18) & \textbf{18.80h}(0.90) & T.O. & 22.55h \\ %
PHP004 & 17.97h(<.01) & \textbf{8.66m}(0.76) & N/A & 77.75m(0.45) & 48.82m(0.21) & N/A & 12.20m(0.97) & 34.19m(<.01) & 9.85m & 12.86m \\ %
PHP009 & 25.55m(<.01) & 7.04m(0.11) & N/A & 50.93m(<.01) & 6.96m(0.89) & N/A & 5.91m(0.14) & 9.05m(<.01) & \textbf{2.81m} & 8.01m \\ %
PNG001 & T.O.(0.02) & 22.73h(0.07) & T.O.(0.02) & T.O.(0.02) & 21.61h(0.11) & 23.24h(0.06) & T.O.(0.02) & T.O.(0.02) & 22.12h & \textbf{18.46h} \\ %
PNG007 & 9.97h(<.01) & 3.86h(<.01) & T.O.(<.01) & 4.07h(<.01) & 1.68h(<.01) & 10.54h(<.01) & 9.88h(<.01) & 46.99m(<.01) & \textbf{8.44m} & 29.83m \\ %
SND017 & 15.02m(0.55) & 17.23m(0.24) & N/A & 31.44m(0.14) & 10.99h(<.01) & 27.68m(0.04) & \textbf{32.40s}(1.00) & 13.18m(0.56) & 13.27m & 15.10m \\ %
SND020 & 37.07m(<.01) & 38.38m(<.01) & N/A & 3.04h(<.01) & 19.92h(<.01) & 68.89m(<.01) & \textbf{11.74m}(0.92) & 27.17m(0.21) & 20.96m & 25.15m \\ %
SQL002 & 67.21m(<.01) & 13.84m(0.14) & N/A & 55.77m(<.01) & \textbf{6.55m}(0.76) & 45.34m(<.01) & 37.15m(<.01) & 9.84m(0.17) & 13.68m & 7.73m \\ %
SQL003 & 21.87h(0.19) & 21.83h(0.19) & N/A & 22.57h(0.17) & T.O.(0.04) & 22.57h(0.17) & T.O.(0.04) & T.O.(0.04) & 23.79h & \textbf{20.12h} \\ %
SQL012 & 20.06h(0.06) & 22.61h(<.01) & N/A & 20.09h(0.05) & 23.88h(<.01) & 16.67h(0.28) & T.O.(<.01) & 17.77h(0.23) & 19.15h & \textbf{15.70h} \\ %
SQL013 & 22.81h(0.44) & 22.93h(0.25) & N/A & 22.91h(0.25) & 23.29h(0.25) & \textbf{18.73h}(0.84) & T.O.(0.08) & 22.15h(0.44) & T.O. & 21.42h \\ %
SQL014 & 3.45h(<.01) & 2.29h(<.01) & N/A & 3.63h(<.01) & 2.17h(0.06) & 3.20h(<.01) & 4.43h(<.01) & 44.51m(0.06) & 25.32m & \textbf{22.26m} \\ %
SQL015 & 23.10h(0.86) & 22.68h(0.86) & N/A & 21.39h(0.94) & \textbf{20.03h}(0.97) & T.O.(1.00) & T.O.(1.00) & T.O.(1.00) & T.O. & T.O. \\ %
SQL020 & 18.33h(<.01) & 22.41h(<.01) & N/A & 20.90h(<.01) & 14.94h(0.01) & 17.85h(<.01) & T.O.(<.01) & 13.02h(0.01) & 15.16h & \textbf{5.96h} \\ %
SSL001 & 6.49h(<.01) & 10.19h(<.01) & 21.26h(<.01) & 9.30h(<.01) & 14.40h(<.01) & 4.98h(0.21) & 18.61h(<.01) & 3.64h(0.03) & 3.61h & \textbf{59.40m} \\ %
SSL020 & 21.59h(<.01) & 21.07h(<.01) & 3.74h(0.03) & 21.03h(<.01) & 22.89h(<.01) & 10.21h(0.01) & 16.66h(<.01) & 6.79h(<.01) & \textbf{1.68h} & 2.41h \\ %
TIF001 & T.O.(0.18) & T.O.(0.18) & T.O.(0.18) & T.O.(0.18) & \textbf{23.32h}(0.56) & T.O.(0.18) & T.O.(0.18) & T.O.(0.18) & 23.91h & T.O. \\ %
TIF002 & T.O.(<.01) & \textbf{9.93h}(0.96) & T.O.(<.01) & 16.09h(0.61) & 11.31h(0.93) & 21.30h(0.05) & 16.73h(0.55) & 14.39h(0.69) & 19.90h & 16.38h \\ %
TIF005 & T.O.(0.18) & 23.60h(0.50) & \textbf{10.45h}(0.99) & 23.85h(0.50) & T.O.(0.18) & T.O.(0.18) & 22.12h(0.56) & 21.97h(0.56) & T.O. & 23.38h \\ %
TIF006 & 13.12h(0.58) & 15.35h(0.45) & \textbf{8.89h}(0.93) & 17.43h(0.28) & 10.09h(0.90) & 17.20h(0.22) & 8.90h(0.90) & 9.32h(0.83) & 14.12h & 17.25h \\ %
TIF008 & T.O.(0.04) & T.O.(0.04) & T.O.(0.04) & 22.26h(0.19) & T.O.(0.04) & 23.33h(0.17) & T.O.(0.04) & T.O.(0.04) & \textbf{22.09h} & T.O. \\ %
TIF009 & 11.63h(<.01) & 15.96h(<.01) & T.O.(<.01) & 23.25h(<.01) & 11.98h(<.01) & 7.94h(<.01) & 6.08h(<.01) & 4.61h(0.01) & \textbf{56.93m} & 8.78h \\ %
TIF014 & 53.89m(<.01) & 92.69m(<.01) & 15.84h(<.01) & 64.99m(<.01) & 81.53m(<.01) & 81.67m(<.01) & 16.86m(0.03) & 28.20m(0.31) & \textbf{8.21m} & 43.59m \\ %
XML001 & T.O.(<.01) & 21.72h(<.01) & 8.86h(<.01) & 20.37h(<.01) & 15.35h(<.01) & 22.12h(<.01) & 13.81h(<.01) & 4.77h(<.01) & 83.88m & \textbf{54.90m} \\ %
XML002 & T.O.(1.00) & T.O.(1.00) & T.O.(1.00) & T.O.(1.00) & 22.11h(0.94) & T.O.(1.00) & T.O.(1.00) & \textbf{21.06h}(0.97) & T.O. & T.O. \\ %
XML003 & 31.68m(0.04) & 22.08m(0.05) & \textbf{2.61m}(0.99) & 1.71h(<.01) & 39.14m(<.01) & 2.84h(<.01) & 55.83m(0.02) & 13.82m(0.66) & 14.56m & 16.06m \\ %
XML006 & T.O.(0.18) & T.O.(0.18) & T.O.(0.18) & 23.18h(0.50) & T.O.(0.18) & T.O.(0.18) & T.O.(0.18) & \textbf{22.10h}(0.56) & 22.36h & T.O. \\ %
XML009 & 42.69m(<.01) & \textbf{6.69m}(0.37) & 13.28h(<.01) & 37.59m(<.01) & 11.95m(0.06) & 1.68h(<.01) & 16.41m(<.01) & 28.22m(0.17) & 8.04m & 9.31m \\ %
XML010 & T.O.(<.01) & 23.18h(<.01) & T.O.(<.01) & T.O.(<.01) & 22.36h(<.01) & T.O.(<.01) & T.O.(<.01) & 18.89h(0.01) & 14.63h & \textbf{8.75h} \\ %
XML012 & T.O.(<.01) & 20.48h(<.01) & T.O.(<.01) & 21.71h(<.01) & 13.93h(0.05) & 22.43h(<.01) & 11.64h(0.09) & 22.21h(<.01) & 19.68h & \textbf{8.82h} \\ %
\bottomrule
\end{tabular}%
\end{table*}

%% file: ablation_table.tex
\begin{table}
\centering
\caption{Ablation Results}
\label{AblationResults}
\resizebox{\columnwidth}{!}{
\setlength\tabcolsep{3pt}
\begin{tblr}{
  cells = {c},
  cell{1}{2} = {c=3}{},
  cell{1}{5} = {c=3}{},
  hlines,
  vlines,
}
       & without False Positives &  &  & with False Positives &  &  \\
Bug ID & \sysname{} & \sysname{}- & \sysname{}* & \sysname{} & \sysname{}- & \sysname{}* \\
CVE-2023-25588 & 11.26h & T.O. & 9.25h & 14.97h & T.O. & 16.91h \\
CVE-2023-25587 & 6.86h & T.O. & 5.62h & 10.44h & T.O. & 11.86h \\
CVE-2022-44408 & 0.31h & 13.14h & 1.20h & 0.59h & 5.58h & 23.50h \\
CVE-2018-13785 & 0.55h & T.O. & 0.92h & 0.42h & T.O. & 0.56h \\
CVE-2013-6954 & 0.57h & 3.50h & 0.37h & 0.46h & 2.74h & 0.32h \\
\end{tblr}
}
\end{table}

%% file: Discussion.tex
In this section, we discuss the scope and limitation of \sysname{}, and the potential enhancements that could be applied to our work in the future to address any limitations.

\subsection{Scope}

\vspace{2pt}\noindent\textbf{Target path-diversity metric}.
Our target path-diversity metric aims to address PFA and PTA, as previously stated. In cases where the bug does not require extra paths for activation and can be triggered through mutations of seeds covering the buggy target location only, extra seeds aimed at enhancing path diversity become redundant. If this information is known in advance, users may choose to exclusively utilize the unbiased energy assignment (i.e., \sysname{}-).

\vspace{2pt}\noindent\textbf{Unbiased energy assignment}.
The scope of unbiased energy assignment is significantly broader by comparison. This method can be applied to nearly all scenarios of directed fuzzing. The scenario in which \sysname{} may become less effective occurs when the buggy target location is shallow and easily reachable, whereas false-positive target locations are deep and challenging to reach. In such instances, the biased energy assignment utilized by other directed fuzzing approaches might coincidentally bias to the correct target.

\subsection{Limitations and Potential Solutions}

\vspace{2pt}\noindent\textbf{Seed explosion}. Similar to other works that propose a more fine-grained coverage metric~\cite{Chen2018AngoraEF,Wang2021ReinforcementLH,Coppik2019MemFuzzUM,Mans2020AnkouGG}, our technique also suffers from problem of seed explosion.
This phenomenon can lead to substantial overhead when the target count reaches into the thousands, given that the cost associated with processing each new seed is significant.
In the current design, each target cluster corresponds to a virgin map that is exactly the same as the primary virgin map. However, as noted in \cite{luo2022selectfuzz}, not all edges are relevant to targets, and these irrelevant edges can be discarded through static analysis. Consequently, their orthogonal approach could be integrated with our work to potentially mitigate the seed explosion problem.

\vspace{2pt}\noindent\textbf{Local optimum}.
In the current design, the critical boundary block requires that the path toward the target should not include any previously covered blocks.
However, this can lead to directed fuzzing getting stuck in a local optimum, since progress toward the target may not necessarily result from mutating the seed that covers the boundary block.
Although the addition of a vector $c$ to seed weight vector as a mitigation strategy is proposed in \cref{EnergyAssignment}, it fails to tackle the root of the issue.
A potential solution could involve relaxing the restriction on the number of non-covered blocks in the path: rather than insisting that no blocks can be covered, permitting a small number, $n$, of blocks to be covered might be a viable approach.

\vspace{2pt}\noindent\textbf{Clustering algorithm}.
Currently, \sysname{} employs a simple and naive clustering algorithm based on a simplified version of association rule mining.
However, more accurate clustering algorithms in the context of association rule mining exist~\cite{Han1997ClusteringBO}. As a future direction, the clustering algorithm could be improved by adopting a more sophisticated approach without sacrificing efficiency.

%% file: Related.tex
\subsection{Improving Coverage Metric}

Numerous studies aim to enhance fuzzing effectiveness by refining the coverage metric. Angora~\cite{Chen2018AngoraEF} considers call context under which each edge is covered to make coverage metric context-sensitive. MemFuzz~\cite{Coppik2019MemFuzzUM} uses address of memory access as extra coverage metric. Ankou~\cite{Mans2020AnkouGG} designs a new fitness function based on distance between execution paths of test cases, and their approach also allows to store extra seeds that do not achieve new coverage. Ijon~\cite{Aschermann2020IjonED} improves coverage metric to explore deep state in specific tested program by taking advantage of the guidance of user-defined annotations marked on important data in the program. AFL-Hier~\cite{Wang2021ReinforcementLH} proposes a multi-level coverage metric and a reinforcement-learning-based hierarchical scheduler in order to handle seed explosion problem introduced by more fine-grained coverage metric.

Some other works also have proposed coverage metric in order to find domain-specific bugs. SlowFuzz~\cite{Petsios2017SlowFuzzAD} uses number of executed instructions as metric to find bugs caused by worst case of algorithm complexity. PerfFuzz~\cite{Lemieux2018PerfFuzzAG} similarly uses maximum count of each program location as storage metric of new seed to also find complexity bugs. MemLock~\cite{Wen2020MEMLOCKMU} proposes memory usage as guidance to store extra seeds in order to find uncontrolled memory consumption bugs. Krace~\cite{Xu2020KraceDR} designs a coverage tracking metric specially designed to find bugs caused by data races in kernel file system.

All of these works have developed new coverage metrics for coverage-guided fuzzing, while in contrast, \sysname{} introduces a new coverage metric specifically for directed fuzzing. To the best of our knowledge, we are the first to primarily focus on this task, utilizing the coverage metric based on multiple virgin maps for targets.
Moreover, some of the mentioned works~\cite{Mans2020AnkouGG,Wang2021ReinforcementLH} devise novel seed scheduling algorithms to address the seed explosion problem caused by their new coverage metrics. Similarly, \sysname{} also proposes a new energy assignment algorithm for directed fuzzing, which not only serves the path diversity but also tackles the seed explosion problem.

\subsection{Directed Fuzzing}

AFLGo~\cite{Bhme2017DirectedGF} firstly proposes directed greybox fuzzing to guide greybox fuzzing toward a set of user-defined targets. It proposes a distance oracle for each seed, which is used for energy assignment to favor seed closer to targets. However, such distance oracle is a very crude metric representing preference of seed with respect to targets. Hawkeye~\cite{Chen2018HawkeyeTA} improves directed fuzzing on top of design of AFLGo based on several desired properties it should hold. Parmesan~\cite{sterlund2020ParmeSanSG} finds targets using information from compiler sanitizer passes, and it also proposes some methods of directed fuzzing such as dynamically constructing ICFG during fuzzing campaign. WindRanger~\cite{Du2022WindrangerAD} improves distance computation from a seed to targets by only considering basic blocks that deviate from path toward targets. However, it does not take multiple targets into account when obtaining such blocks. In addition, their definition of deviation block is local and is based on only one seed, while our critical blocks are global and take all seeds in corpus into account. LeoFuzz~\cite{Liang2022MultipleTD} is the first work that realizes the problem of using harmonic average distance of all targets, and it proposes a method based on target sequence in order to solve it. FishFuzz~\cite{Zheng2022FishFuzzTL} also addresses such problem by using a distance vector for each seed instead of an average distance. By contrast, energy assignment of \sysname{} not only solves such problem in a more fine-grained way, but also serves for our path diversity component. Finally, CAFL~\cite{Lee2021ConstraintguidedDG} takes sequence of locations with constraints as target, but such sequence requires more manual effort than other directed fuzzers.

There are other works that try to improve performance of directed fuzzing by trimming redundant parts considered irrelevant to targets in fuzzing campaign. FuzzGuard~\cite{Zong2020FuzzGuardFO} uses deep learning to filter out test case considered to be unhelpful for reaching target, so it is not executed by tested program beforehand. BEACON~\cite{Huang2022BEACONDG} applies static analysis to terminate program early if current program state is guaranteed to be unable to reach any targets. SieveFuzz~\cite{Srivastava2022OneFD} also terminates the program early by restricting fuzzing to search space guaranteed relevant to reaching targets. SelectFuzz~\cite{luo2022selectfuzz} selectively instruments and explores only target-relevant code, and it also proposes a novel distance metric from a basic block to target based on multi-path reaching probability. These works about trimming are considered orthogonal to our approach.

%% file: Conclusion.tex
We proposed \sysname{}, which includes target path-diversity metric and unbiased energy assignment.
Through evaluation, we demonstrated the effectiveness of \sysname{} in triggering known and zero-day bugs.

%% file: Appendix.tex
\input{MathOpt.tex}

\section{Target Weight Calculation} \label{TargetWeightCalculation}

At compile time, \sysname{} can be provided with a set of target locations with their corresponding weight. Target location is a pair of file name and line number, usually in format {\tt file.c:123}. Target basic block, on the other hand, is the basic block in tested program that contains at least one target location. We show how we compute weight of target block from weight of target location in this section. Each target location can be contained by one or more target basic blocks, and each target block can also contain one or more target locations. With such information, we distribute weight from target locations to target blocks using distribution matrix. Each column of the matrix represent a target location, and each row of the matrix represent a target block. We define matrix $\Phi$ as follows:

$$
{\Phi}_{b, l} =
\begin{cases}
 \frac{1}{N_l} & \text{if target block $b$ contains target location $l$} \\
 0 & \text{otherwise}
\end{cases}
$$

$N_l$ is the total number of target blocks that contain target location $l$, and we should note that each column of this matrix sums to $1$. Finally, we can multiply $\Phi$ by weight vector of target locations to obtain weight of each target block. At this point, we can discard these target locations, and unless otherwise specified, we use target to denote target basic block instead of target location in this paper.

\input{CVE.tex}

\section{Static Distance Computation} \label{DistanceComputation}

We have developed an efficient method for computing the distance from each block to each target in $O(|T||E| + |T||V|log|V|)$, where $V$ and $E$ represent vertices and edges of the ICFG, and $T \subset V$ denotes the target blocks. We firstly define the weight of each edge from vertex $v_{src}$ to vertex $v_{dst}$ as follows:

$$
W(v_{src}, v_{dst}) =
\begin{cases}
 {log}_2(N_{out}(v_{src})) & \text{for control-flow edge} \\
 0 & \text{for call edge}
\end{cases}
$$

The term $N_{out}(v_{src})$ represents the number of control-flow outgoing edges of vertex $v_{src}$. The rationale behind this weight definition is to assign higher weight to edges that are less likely to be traversed after executing the basic block $v_{src}$. As a result, the weight of both call edges and unconditional control-flow edges should be $0$, while the weight of conditional control-flow edges grows logarithmically with respect to the number of outgoing edges.

After obtaining the weight for each edge, we can define the distance from block $v$ to target $t$ as the sum of weight along the shortest path from $v$ to $t$. This can be computed using Dijkstra's algorithm~\cite{Dijkstra1959ANO}. A naive approach involves applying Dijkstra's algorithm for each $v \in V$, as done by AFLGo~\cite{Bhme2017DirectedGF}. However, this method has a time complexity of $O(|V||E| + |V|^2log|V|)$, which can be very slow for large ICFGs. Instead, we compute this distance by \emph{reversing each edge} of the ICFG and applying Dijkstra's algorithm for each $t \in T$, since the shortest distance from $v$ to $t$ in the ICFG is equal to the shortest distance from $t$ to $v$ in the transposed ICFG.
As the set of targets $T$ is usually much smaller than the set of basic blocks $V$ in directed fuzzing, this method is significantly more efficient than the naive one.
We describe this method in Algorithm \ref{GetDistance}.

\begin{algorithm}[hbt!]
\caption{Distance Computation} \label{GetDistance}
\begin{algorithmic}[1]
\REQUIRE {\emph{ICFG}}

\STATE ICFG' = transpose(ICFG) \\
\STATE distance: $V \times T \rightarrow \mathbb{R}$
\FOR {$t \in T$}
    \FOR {(b, d) $\in$ Dijkstra(ICFG', t)}
        \STATE distance[b, t] = d
    \ENDFOR
\ENDFOR

\RETURN distance

\end{algorithmic}
\end{algorithm}

\section{Implementation}

\input{Implementation.tex}

\section{Quality of Extra Seeds} \label{QualityExtra}

In this appendix section, we address RQ5 by counting the number of unique crashes that are mutated from the extra seeds generated by our target path-diversity metric and calculating its ratio over the total number of crashes.
We continued to use the results from the Magma benchmark~\cite{Hazimeh2020MagmaAG} and enabled the fatal canary feature in Magma, causing any bug trigger to result in a crash. For each seed generated during the fuzzing campaign, \sysname{} annotates its filename with the factor contributing to its storage.
Specifically, if a seed covers any new bits in the primary virgin map, \sysname{} marks a \texttt{cov} flag in its filename, and similarly adds a \texttt{div} flag if any new bits are covered in the target virgin map. Moreover, each unique crash is marked with the seed ID from which the unique crash is mutated, also known as the source seed.
Using this information, we can track the flags of the source seed for each unique crash. We consider a unique crash to be mutated from the extra seed only if its source seed contains solely the \texttt{div} flag. We denote such unique crash as the diversity crash. The results are presented in Table \ref{UniqueCrashes}.
We observe that for the majority of programs, more than half of the unique crashes are mutated from our extra seeds, demonstrating the ability of these extra seeds to discover vulnerabilities.

\begin{table}[]
\caption{Statistics about Unique Crashes}
\label{UniqueCrashes}
\resizebox{\columnwidth}{!}{%
\begin{tabular}{cccc}
\toprule
Program & No. Diversity Crashes & No. Total Crashes & Ratio \\
\midrule
libpng & 1336 & 1437 & 92.97\% \\
libsndfile & 97 & 169 & 57.40\% \\
libtiff & 4485 & 5476 & 81.90\% \\
libxml2 & 3711 & 4448 & 83.43\% \\
lua & 430 & 584 & 73.63\% \\
openssl & 31 & 38 & 81.58\% \\
php & 943 & 1084 & 86.99\% \\
poppler & 753 & 1076 & 69.98\% \\
sqlite3 & 675 & 1395 & 48.39\% \\
\bottomrule
\end{tabular}%
}
\end{table}

\input{eval_table2.tex}

\input{ObtainCluster.tex}

\section{Additional Experiments} \label{AdditionalExperiments}

In addition to the Magma benchmark, we conducted further experiments to evaluate the effectiveness of \sysname{} compared to other fuzzers using a set of vulnerabilities used in the evaluations of previous directed fuzzing works~\cite{Kim2023DAFLDG, Huang2022BEACONDG}.
All fuzzers in \cref{MagmaEvaluation} are included in this experiment except for \sysname{}-{}-, the ablated version of \sysname{} that does not contain the target path-diversity metric and critical blocks for evaluating the effectiveness of critical blocks in our energy assignment method, a task we have already accomplished using the Magma benchmark. However, we still include \sysname{}- for its descent ability to find the vulnerabilities that do not require path diversity to trigger.

The experimental setup was identical to that described in \cref{MagmaEvaluation}: each fuzzing campaign was conducted 10 times, with each lasting 24 hours.
However, we directly used the buggy version of the tested program instead of reintroducing old bugs into newer versions. Additionally, we lacked the automatic vulnerability trigger time recording feature provided by Magma.
To overcome this limitation, each fuzzer was configured to record the time when a crashing input is saved. After each fuzzing campaign ends, we executed the files in the crash directory and identified the specific vulnerability that causes each crash, based on the execution's output. This allows us to determine the triggering time of a vulnerability in each fuzzing campaign.
For the initial fuzzing setup, we utilized the same seed corpus as the previous work~\cite{Kim2023DAFLDG}.
However, the target locations were different: we adjusted the target locations to the root-cause locations instead of the vulnerability-trigger locations commonly used in previous works; besides, similar to \cref{MagmaEvaluation}, we utilized multiple-target settings to find multiple vulnerabilities simultaneously. These two differences in target locations align with the two main issues addressed in our work.

The results are presented in Table~\ref{NewBugResults}, formatted similarly to Table \ref{MagmaBenchmarkResults}.
Compared to AFL++, AFLGo, Parmesan, FishFuzz, Hawkeye, WindRanger, and MOpt, \sysname{}(-) achieves speedups of 164\%, 77\%, 819\%, 319\%, 136\%, 118\%, and 194\% respectively.
We observe the deviation of the results from those obtained from Magma: AFLGo instead of MOpt becomes the best performer among all non-\sysname{} fuzzers, and the effectiveness of Parmesan drops significantly (even when excluding unrunnable campaigns due to its buggy implementation).
Furthermore, our results also differ from those of previous studies, as AFLGo outperforms the works that followed it. We believe this inconsistency arises from our different configuration of the target locations mentioned earlier.
Nevertheless, \sysname{} still outperforms the other tested fuzzers, demonstrating its superiority across a diverse set of bugs without any overfitting.

\input{MathOptAlgo.tex}

%% file: MathOpt.tex
\section{Mathematical Optimization Algorithm} \label{MathematicalOptimizationAlgorithm}

We can state the problem as follows. We are given with a ratio vector $r \in \mathbb{R}^n$ that satisfies $\sum_{i=1}^n r_i = 1$, a vector $b \in \mathbb{R}^n$ and a scalar value $E \in \mathbb{R}$, and these values should all be non-negative. We want to find a non-negative vector $x \in \mathbb{R}^n$ such that $\sum_{i=1}^n x_i = E$, so that distribution of $x + b$ can approach ratio vector $r$ as much as possible. A naive algorithm is to convert the problem into a convex optimization problem by minimizing cross entropy loss between normalized $x + b$ and $r$ under certain constraints. However, if $n$ is large, state-of-art solvers are not able to get the optimal solution in feasible time limit. Instead, we develop an algorithm with worst time complexity $O(n^2)$ that guarantees the optimal solution.

The key intuition of the algorithm is to calculate the sum of total value after assignment by $E' = E + \sum_{i=1}^n b_i$, with such value we can calculate the most desirable result for $x+b$ by $d = {E'}r$. In ideal situation where $d_i \geq b_i$ for $i \in \{1, ..., n\}$, we can achieve such most desirable result by $x = d - b$.

However, in reality, there must be some entries where $d_i < b_i$. Since elements of $x$ cannot be negative, we cannot do $x = d - b$. The idea is that for all $i$ such that $d_i < b_i$, we can know that we must let $x_i = 0$. Therefore, we can remove all such entries to create another sub-problem, and we do this recursively until $d_i \geq b_i$ always holds. The algorithm is detailed in Algorithm \ref{MathOpt}.

%% file: CVE.tex
\section{CVE-2023-25584} \label{CVE202384}

We demonstrate how we were able to find CVE-2023-25584, a group of out-of-bounds read vulnerabilities in \texttt{binutils}, using \sysname{}.
The first vulnerability in this group we have discovered occurs at the \texttt{parse\_module} function in the \texttt{vms-alpha.c} file.
This bug was discovered by targeting recently changed code locations, and indeed, the file had been recently modified to fix some bugs. The root cause of this bug, although not directly related to these changed code locations, could also be tested when fuzzing these targets, which ultimately led to its discovery.

We subsequently analyzed this vulnerability and discovered that numerous nearby locations appeared to exhibit a similar issue due to the access of the same pointer, \texttt{ptr}. We wrote a simple static taint analysis query using CodeQL~\cite{Moor2007KeynoteA,CodeQLDataFlow} to collect all locations that access \texttt{ptr}. We then initiated several concurrent fuzzing campaigns using these locations as targets. Consequently, a significant number of vulnerabilities were uncovered within 24 hours.

For these fuzzing campaigns, the initial corpus was unable to cover any blocks that could reach any targets in the ICFG due to the indirect call.
This resulted in an empty set of critical block, prompting \sysname{} to first run in coverage-guided mode.

As the fuzzing campaigns progress, coverage-guided fuzzing manages to find a seed that covers some of these blocks, generating critical blocks and prompting \sysname{} to transition into directed fuzzing mode. However, certain constraints along the path to the target sites are difficult to solve, preventing other fuzzers from hitting these targets and causing these vulnerabilities to remain undetected by OSS-Fuzz. Thanks to our energy assignment algorithm, \sysname{} assigns a significant amount of energy to fuzz this seed, effectively passing these constraints and subsequently covering the targets. By fuzzing these targets, a group of out-of-bounds read vulnerabilities has been identified.

We list all vulnerability IDs of this CVE as follows: 29848, 29882, 29873, 29874, 29875, 29876, 29877, 29878, 29879, 29880, 29881, 29883, 29884, 29885, 29886, 29887, 29888, 29889, 29890, 29891.

%% file: Implementation.tex
We have implemented \sysname{}, a prototype based on the ideas presented, using the source code from AFL++~\cite{Fioraldi2020AFLC} and AFLGo~\cite{Bhme2017DirectedGF}, totaling over 5,000 lines of C/C++ code. In this section, we will discuss some implementation details.

\vspace{2pt}\noindent\textbf{Static analysis and instrumentation}.
The compilation stage of \sysname{} is built upon the LLVM~\cite{Lattner2004LLVMAC} Link Time Optimization (LTO) pass~\cite{LLVMLTO} by modifying the AFL++ LTO compiler~\cite{AFLLTO}.
First, we identify all target basic blocks. Then, using information from LLVM, we construct the ICFG of the PUT. With the ICFG, we gather all basic blocks that can reach at least one of the target blocks, including the target blocks themselves, and calculate their distances to each target block using the method described in Appendix \ref{DistanceComputation}. Each of these basic blocks is assigned a unique ID. Moreover, to obtain critical blocks at runtime, we record the subgraph of the ICFG with these blocks as vertices. These blocks are instrumented with a function at the beginning, which will be called with the block ID as an argument when the block is executed.

\vspace{2pt}\noindent\textbf{Virgin map comparison}. 
In AFL++, when comparing the execution trace with the virgin map, it avoids inefficiently comparing each bit individually. Instead, it compares 64 bits at a time by casting pointers into 64-bit arrays. For each 64-bit value in the execution trace and virgin map, AFL++ calculates their bitwise {\tt AND} value, and a non-zero result indicates new bit coverage by the execution trace.
We employ the same technique for efficient comparisons with multiple virgin maps by calculating the bitwise {\tt AND} value between the execution trace and each virgin map.
Each non-zero {\tt AND} value at each location for each virgin map is also recorded and used to obtain $data$ in Algorithm \ref{SaveIfInterestingAFLRun}.

\vspace{2pt}\noindent\textbf{Target clusters}. We assign each cluster an ID, and each cluster contains a set of targets recorded using an array indexed by the cluster ID. In addition to recording the set of targets in each cluster, we also need to map each target to its respective cluster in order to efficiently implement \texttt{get\_cluster} function in Algorithm \ref{SelectVirginMaps}. This is achieved using hash map~\cite{RobinHoodHash}.

\vspace{2pt}\noindent\textbf{Association rule mining}. We implement \texttt{support\_count} in Algorithm \ref{UpdateSupportCount} using two data structures: one for counting 1-itemsets and another for counting 2-itemsets. We record 1-itemset support counts using an array indexed by cluster ID, while 2-itemset support counts are recorded using hash map~\cite{RobinHoodHash}, utilizing order-insensitive pairs as keys.

\vspace{2pt}\noindent\textbf{Covered blocks}.
For efficiency, we only consider covered blocks that can reach at least one target in the ICFG, each of which is assigned an ID during static analysis.
To trace the set of blocks covered by the execution of a test case, we use a shared memory with $n$ bits, where $n$ is the number of blocks assigned with an ID. Each bit in memory represents a distinct block.
In the instrumented function for each block, the block ID is obtained from the argument, allowing instrumented code to set the corresponding bit in the shared memory to $1$ to indicate that the block is covered during execution.
Furthermore, similar to the virgin map, we also have a virgin block bitmap with $n$ bits to record each block covered by the corpus, enabling \sysname{} to determine whether a new seed covers any new blocks.

\vspace{2pt}\noindent\textbf{Energy assignment}. The energy assignment algorithm discussed in \cref{EnergyAssignment} involves matrix multiplication. However, since the related matrices are typically sparse, computing matrix multiplication explicitly can be inefficient. To address this, we represent each weight vector as a hash map~\cite{RobinHoodHash} and use the relevant information to directly distribute weight from one hash map to another.

%% file: eval_table2.tex
\begin{table*}[h]
\caption{Additional Results}
\label{NewBugResults}
\footnotesize
\centering
\setlength\tabcolsep{3pt}
\begin{tabular}{cccccccccccc}
\toprule
Program & CVE ID & AFL++ & AFLGo & Parmesan & FishFuzz & Hawkeye & WindRanger & MOpt & \sysname{} & \sysname{}- \\
\midrule

\multirow{5}{*}{cxxfilt} & 2016-4487 & 15.37m(0.01) & 13.90m(0.08) & N/A & 18.27m(<.01) & 33.01m(0.04) & 26.54m(<.01) & 51.80m(<.01) & \textbf{5.31m} & 6.70m \\
 & 2016-4489 & 37.33m(<.01) & 37.55m(<.01) & N/A & 25.28m(<.01) & 15.96m(<.01) & 31.79m(<.01) & 68.94m(<.01) & 3.87m & \textbf{2.88m} \\
 & 2016-4490 & 6.98m(<.01) & 9.22m(<.01) & N/A & 9.25m(<.01) & 3.78m(<.01) & 8.31m(<.01) & 19.82m(<.01) & \textbf{52.20s} & 84.20s \\
 & 2016-4491 & 21.90h(0.86) & \textbf{17.29h}(0.97) & N/A & 18.45h(0.97) & 19.91h(0.97) & 22.50h(0.86) & 23.60h(0.86) & T.O. & T.O. \\
 & 2016-4492 & 41.92m(0.17) & 38.44m(0.09) & N/A & 46.37m(0.09) & 2.94h(0.04) & 77.24m(0.07) & 97.45m(<.01) & \textbf{30.30m} & 59.77m \\ \hline
strip & 2017-7303 & 2.95m(0.96) & \textbf{2.88m}(0.95) & 8.45h(<.01) & 36.72m(0.34) & 6.70m(0.76) & 74.21m(<.01) & 4.47m(0.24) & 7.39m & 23.49m \\ \hline
\multirow{3}{*}{objcopy} & 2017-8393 & \textbf{27.67m}(0.17) & 36.43m(0.21) & 13.30h(<.01) & 3.55h(0.02) & 4.48h(0.02) & 1.79h(0.09) & 28.95m(0.43) & 62.99m & 36.18m \\
 & 2017-8394 & 26.95m(0.92) & 14.34m(0.94) & 5.15h(0.48) & 6.17h(<.01) & 3.00h(0.19) & 9.10m(0.98) & \textbf{8.92m}(1.00) & 67.11m & 1.97h \\
 & 2017-8395 & 6.64m(0.09) & 1.88m(0.29) & 30.34m(<.01) & 3.71h(<.01) & \textbf{94.10s}(0.63) & 2.03m(0.09) & 2.94m(<.01) & 1.73m & 2.39m \\ \hline
\multirow{4}{*}{objdump} & 2017-8392 & 97.68m(<.01) & 6.31m(<.01) & 82.40s(<.01) & 8.14h(<.01) & 3.73m(<.01) & 2.86m(<.01) & 8.67m(<.01) & \textbf{16.50s} & 33.60s \\
 & 2017-8396 & \textbf{6.42h}(1.00) & 22.58h(0.56) & 19.30h(0.79) & T.O.(0.18) & T.O.(0.18) & T.O.(0.18) & 22.71h(0.56) & T.O. & 23.89h \\
 & 2017-8397 & 19.75h(0.03) & 21.99h(<.01) & 21.62h(<.01) & 21.15h(<.01) & 21.44h(<.01) & 18.66h(0.05) & 20.41h(0.02) & 13.61h & \textbf{12.51h} \\
 & 2017-8398 & 1.81h(0.50) & 6.88m(0.26) & 7.53h(<.01) & 8.44h(<.01) & 20.14m(0.01) & \textbf{4.03m}(0.87) & 9.10m(0.07) & 7.56m & 13.10m \\ \hline
\multirow{15}{*}{swftophp} & 2016-9827 & 4.01m(<.01) & 1.67m(0.02) & N/A & 86.70s(0.04) & 62.10s(0.37) & 89.00s(0.04) & 86.50s(0.03) & \textbf{61.20s} & 89.10s \\
 & 2016-9829 & 33.59m(<.01) & 9.14m(0.03) & N/A & 8.30m(0.57) & 7.96m(0.71) & 11.39m(0.26) & \textbf{4.95m}(0.71) & 7.60m & 6.55m \\
 & 2016-9831 & 14.52m(<.01) & 8.57m(0.03) & N/A & 11.23m(0.01) & 6.90m(0.12) & 7.89m(0.11) & 10.86m(<.01) & \textbf{4.31m} & 6.84m \\
 & 2017-11728 & 66.83m(0.01) & 24.57m(0.14) & N/A & 45.58m(0.02) & \textbf{21.74m}(0.04) & 78.32m(<.01) & 55.13m(0.01) & 2.81h & 46.11m \\
 & 2017-11729 & 4.54m(<.01) & 2.74m(0.10) & N/A & 7.00m(0.05) & 2.87m(0.26) & 12.55m(<.01) & 19.49m(<.01) & \textbf{1.80m} & 4.02m \\
 & 2017-9988 & 7.17h(<.01) & 59.18m(0.26) & N/A & 2.32h(0.19) & 3.82h(0.21) & 17.66h(<.01) & 53.43m(0.15) & \textbf{52.56m} & 5.11h \\
 & 2017-7578 & 12.48m(<.01) & 7.41m(<.01) & N/A & 3.30m(0.01) & 7.59m(0.02) & 5.82m(0.02) & 6.55m(<.01) & \textbf{1.70m} & 1.95m \\
 & 2018-11095 & 1.72h(<.01) & 83.75m(<.01) & N/A & 8.13h(<.01) & 5.52h(<.01) & 80.06m(<.01) & 4.25h(<.01) & 33.31m & \textbf{7.58m} \\
 & 2018-11225 & 14.27h(0.32) & 16.72h(0.05) & N/A & 19.95h(<.01) & T.O.(<.01) & 13.65h(0.45) & 18.61h(0.02) & 20.66h & \textbf{9.77h} \\
 & 2018-20427 & 11.94h(0.28) & 12.57h(0.09) & N/A & 14.46h(0.04) & 14.52h(0.02) & 7.37h(0.79) & 9.22h(0.31) & 12.74h & \textbf{6.28h} \\
 & 2018-7868 & 3.25h(0.83) & 3.19h(0.74) & N/A & 8.59h(0.08) & 11.75h(<.01) & \textbf{3.02h}(0.81) & 5.93h(0.11) & 7.42h & 3.48h \\
 & 2018-8807 & 2.27h(0.26) & 2.47h(0.52) & N/A & 7.37h(<.01) & 3.79h(0.14) & 2.40h(0.66) & 4.48h(<.01) & 5.69h & \textbf{2.06h} \\
 & 2019-12982 & 8.80h(0.42) & 14.52h(0.20) & N/A & 15.08h(0.01) & 20.80h(<.01) & \textbf{6.21h}(0.93) & 13.72h(0.07) & 21.13h & 6.93h \\
 & 2020-6628 & 2.69h(0.08) & 1.93h(0.71) & N/A & 7.64h(<.01) & 4.56h(0.08) & \textbf{83.77m}(0.96) & 4.25h(<.01) & 4.08h & 1.96h \\
 & 2019-9114 & \textbf{5.30h}(0.94) & 19.03h(<.01) & N/A & 14.22h(0.01) & 19.23h(<.01) & 9.56h(0.40) & 17.01h(0.02) & 13.63h & 7.30h \\ \hline
\multirow{3}{*}{xmllint} & 2017-9047 & 23.46h(<.01) & 22.00h(<.01) & N/A & 22.49h(<.01) & T.O.(<.01) & 18.49h(0.02) & T.O.(<.01) & 23.26h & \textbf{12.74h} \\
 & 2017-5969 & 30.08m(<.01) & 50.30s(1.00) & N/A & 28.10m(<.01) & \textbf{28.30s}(0.99) & 57.05m(<.01) & T.O.(<.01) & 3.17m & 2.36m \\
 & 2017-9048 & 23.68h(0.06) & T.O.(0.02) & N/A & T.O.(0.02) & T.O.(0.02) & 23.16h(0.07) & T.O.(0.02) & 22.75h & \textbf{20.99h} \\ \hline
\multirow{2}{*}{lrzip} & 2017-8846 & 18.89h(<.01) & T.O.(<.01) & N/A & 2.15h(0.69) & 4.77h(0.40) & 2.96h(0.89) & 21.61h(<.01) & 7.43h & \textbf{56.77m} \\
 & 2018-11496 & 14.10s(0.66) & 47.10s(0.09) & N/A & 23.20s(0.83) & 35.60s(0.58) & \textbf{10.60s}(0.85) & 47.50s(0.11) & 24.60s & 51.50s \\
\bottomrule
\end{tabular}%
\end{table*}

%% file: ObtainCluster.tex
\section{Obtaining Clusters} \label{ObtainingClusters}

Algorithm \ref{SelectVirginMaps} outlines the process of selecting the set of clusters from targets covered by the execution of a test case.  First, we insert primary cluster corresponding to the primary virgin map. Then, for each target covered by the test case, we find its cluster using $get\_cluster$, which maps each target to the cluster containing it, and insert the returned cluster into $clusters$. Since targets are now clustered and different targets can return the same cluster from $get\_cluster$, we use a unique set as the data type for $clusters$ to prevent duplication of elements.

\begin{algorithm}[hbt!]
\footnotesize
\caption{Obtaining Set of Clusters} \label{SelectVirginMaps}
\begin{algorithmic}[1]
\REQUIRE {\emph{covered\_targets} and \emph{primary\_cluster}}

\STATE clusters = unique\_set() \\
\STATE clusters.insert(primary\_cluster)
\FOR {target $\in$ covered\_targets}
    \STATE clusters.insert(get\_cluster(target))
\ENDFOR
\RETURN clusters

\end{algorithmic}
\end{algorithm}

%% file: MathOptAlgo.tex
\newpage
\begin{algorithm}
\caption{approach\_ratio} \label{MathOpt}
\begin{algorithmic}[1]
\REQUIRE {$E \in \mathbb{R}$, $b \in \mathbb{R}^n$ and $r \in \mathbb{R}^n$}

\STATE $d$ := $(E + \sum_{i=1}^n b_i)r$ \\
\STATE remove := set()
\FOR {i := 1 \TO n}
    \IF {$d_i < b_i$}
        \STATE remove.insert(i)
    \ENDIF
\ENDFOR

\STATE $x \in \mathbb{R}^n$
\IF {$|remove| = 0$}
    \FOR {i := 1 \TO n}
        \STATE $x_i$ := $d_i - b_i$
    \ENDFOR
    \RETURN x
\ENDIF

\STATE $x' \in \mathbb{R}^{n-|remove|}$, $b' \in \mathbb{R}^{n-|remove|}$, $r' \in \mathbb{R}^{n-|remove|}$ \\
\STATE j := 1
\FOR {i := 1 \TO n}
    \IF {i $\notin$ remove}
        \STATE ${b'}_j$ := $b_i$, ${r'}_j$ := $r_i$
        \STATE j := j + 1
    \ENDIF
\ENDFOR

\STATE $x'$ := approach\_ratio($E$, $b'$, $r'$)

\STATE j := 1
\FOR {i := 1 \TO n}
    \IF {i $\in$ remove}
        \STATE $x_i$ := 0
    \ELSE
        \STATE $x_i$ := ${x'}_j$
        \STATE j := j + 1
    \ENDIF
\ENDFOR

\RETURN x

\end{algorithmic}
\end{algorithm}